\documentclass[a4paper,12pt]{article}
\usepackage{jcappub} 
\usepackage{cite}
\usepackage{graphicx}
\usepackage{enumerate}
\usepackage{hyperref}
\usepackage{cancel}
\usepackage{color}
\usepackage{graphicx}
\usepackage{amsmath}
\usepackage{amsfonts}
\usepackage{amssymb}
\usepackage{rotating}
\usepackage{booktabs}
\usepackage{xcolor}
\usepackage{soul}

\def\comment#1{}

\title{\boldmath 
Massive particle pair production and oscillation in Friedman Universe:
reheating energy and entropy, and cold dark matter
}

\author{She-Sheng Xue}

\affiliation{ICRANet Piazzale della Repubblica, 10 -65122, Pescara, Italy
\\ Physics Department, Sapienza University of Rome, 
Rome, Italy\\INFN, Sezione di Perugia, 
Perugia, Italy
\\ICTP-AP, University of Chinese Academy of Sciences, Beijing, China
}

\emailAdd{xue@icra.it and shesheng.xue@gmail.com} 

\abstract{
Suppose that the early Universe starts with a cosmological $\Lambda$-term originating from quantum spacetime at the Planck scale. Dark energy drives inflation and reheating by reducing its value for massive particle-antiparticle pairs production and oscillation, resulting in a holographic and massive pair plasma state. The back-and-forth reaction of dark energy and massive pairs slows inflation to its end and starts reheating by rapidly producing stable and unstable pairs. We introduce the Boltzmann-type rate equation describing the back-and-forth reaction. It forms a close set with Friedman equations and reheating equations for unstable pairs decay to relativistic particles. 
The numerical solutions show preheating, massive pairs dominated and genuine reheating episodes. We obtain the reheating temperature and entropy 
in terms of the tensor-to-scalar ratio 
$0 < r < 0.047$ consistently with observations. Stable massive pairs represent cold dark matter particles and weakly interact with dark energy. The resultant cold dark matter abundance $\Omega_c\sim 10^{-1}$ is about a constant in time.
}

\begin{document}
\maketitle
\flushbottom

\section{\bf Introduction}\label{introduction}
In the standard model of modern cosmology ($\Lambda$CDM), the cosmological constant $\Lambda$, dark matter, inflation, reheating and coincidence problem have been long-standing basic issues for decades. The inflation \cite{Starobinsky1980, Guth1981, Linde:1981mu, Mukhanov:1982nu, Albrecht1982, Linde1983, Kallosh:2021mnu}
reheating \cite{Kofman1994, Kofman1997, Shtanov1995, Bassett1998, Tsujikawa1999, Podolsky2002, Allahverdi2010, Amin2012, Amin2014, Adshead2020} are fundamental processes. The latter transition the Universe from the cold and massive state left by inflation to the hot Big Bang, and then the standard cosmology follows. The evolution follows the Friedman equations of cosmological $\Lambda$, matter and radiation energy densities. The cosmological $\Lambda$ and massive particle origin are still mysteries. One calls them ``dark energy" and ``cold dark matter".
Moreover, their properties and interactions in the Universe's evolution are also in question. Why their present values are coincidentally in the same order of magnitude?

To get an insight into these issues, people have been intensively studying the gravitational particle production in Friedman Universe for decades \cite{PhysRevLett.21.562, PhysRevD.3.346,   PhysRev.183.1057, Zeldovich1971, Parker1973, Mamaev1976, Zeldovich1977, Starobinsky:1979ty, Birrell1984, Mottola1985,Habib2000, Anderson2014,Anderson2014a, Landete2014}. Based on the adiabatic and non-back-reaction approximation for a slowly time-varying Hubble function $H$, one adopted the semi-classical approaches to calculating the particle production rate. It is exponentially suppressed $e^{-M/H}$ for massive particles $M\gg H$ since the classical Universe evolution time scale ${\mathcal O}(1/H)$ is much larger than the quantum time scale ${\mathcal O}(1/M)$ of particle production. However, the non-adiabatic back-reactions of massive particle productions on the Hubble function can be large. One has to take them into account. People have made many efforts 
\cite{Parker1973, 
Starobinsky1982, Ford1987,
Kolb1996,
Chung2001,
Chung2019, Ema2018, 
Xue2019, Xue2020, Xue2021} to study non-adiabatic back-reaction and 
understand massive particle productions without exponential suppression. 
To properly include the back-reaction of particle production on Universe evolution, one should separate fast components ${\mathcal O}(1/M)$ from slow components ${\mathcal O}(1/H)$ in the Friedman equation.  
Here, the fast components represent fluctuating gravitational and particle fields 
at the time scale ${\mathcal O}(1/M)$. The slow components represent slowly varying background fields and particle densities 
at the time scale ${\mathcal O}(1/H)$. 

In Ref.~\cite{Xue2021}, we assume the inflation epoch starts when the dark-energy density $\rho_{_\Lambda}=\Lambda/(8\pi G)$ is the order of the Planck scale, dominates the Hubble function $H^2\approx 8\pi G\rho_{_\Lambda}/3$ and the matter density is negligibly small. We study the Universe undergoes a $\Lambda$-driven inflation that slowly slows down to its end by producing heavy particles of mass $M\gg H$. Inflation is a semi-classical dynamics of the time scale ${\mathcal O}(1/H)$, while massive particle production is a quantum-field dynamics of the time scale ${\mathcal O}(1/M)$. We investigate how the massive particle production density $\rho^H_{_M}$ back reacts $\rho_{_\Lambda}$ 
by separating the fast and slow components in the Friedman equation. Analysis shows $H$ and $\rho_{_\Lambda}$ slowly decrease, as $\rho^H_{_M}$ slowly increases in time, namely, $\rho_{_\Lambda}$ slowly converts to $\rho^H_{_M}$. It gives the quasi-de Sitter phase (slow-rolling dynamics) for inflation. The final results are consistent with observations. Here we turn to discuss the reheating epoch and possibly explain reheating energy and entropy, as well as cold dark matter, in comparison with current observations. 

We review the previous results of the fast and slow components' separation in Sec.~\ref{sepra}, quantum massive pair production and oscillation in Sec.~\ref{qppo}, 
and a massive pair plasma state in Sec.~\ref{mpp}. We present in Secs.~\ref{mppa} and \ref{infend} the new studies of a complete set of differential equations and initial conditions for the reheating epoch. We numerically solve these equations and compare the results with observations in Secs.~\ref{reheatingdetails} and \ref{obsm}. We present preliminary discussions on stable massive as cold dark matter candidate in Sec.~\ref{coldm}. $G=M^{-2}_{\rm pl}$ is the Newton constant, $M_{\rm pl}$ is the Planck scale and
reduced Planck scale $m_{\rm pl}\equiv (8\pi)^{-1/2} M_{\rm pl}=2.43\times 10^{18} $GeV.

\section{Slow adiabatic 
and fast non-adiabatic components}\label{sepra}

We discuss such fast and slow separation in the $\tilde\Lambda$CDM scenario, where a time-varying cosmological $\tilde\Lambda$ term 
in the Friedman equation represents such interacting dark energy. The Friedman equations for a flat Universe are \cite{Xue2015}
\begin{eqnarray}
H^2=\frac{8\pi G}{3}\rho;\quad \dot H=-\frac{8\pi G}{2}(\rho + p),\label{friedman}
\end{eqnarray}
where energy density $\rho\equiv \rho_{_M}+\rho_{_R}+\rho_{_\Lambda}$ and pressure 
$p\equiv p_{_M}+p_{_R}+p_{_\Lambda}$. 
The second Equation of (\ref{friedman}) is the generalised 
conservation law (Bianchi identity) for including time-varying cosmological term
$\rho_{_\Lambda}(t)\equiv \tilde\Lambda/(8\pi G)$. It reduces to the usual Equation 
$\dot \rho_{_M} + (1+\omega_{_M})H\rho_{_M} +\dot \rho_{_R} + (1+\omega_{_R})H\rho_{_R}=0$ for 
time-constant $\rho_{_\Lambda}$.  
The second Equation of (\ref{friedman}) shows that $\dot H <0$ and $H$ decreases in time, 
due to the matter's gravitational attractive nature. 

\comment{We adopt two independent equations of Friedmann and energy conservation law 
to completely determine 
the cosmological energy density $\rho_{_{_\Lambda}}$ and  
inflation rate $H$ \cite{xue2020MPLA},
\begin{eqnarray}
&& H^2 = 	\frac{8\pi G}{3}( \rho_{_\Lambda}  + \rho_{_M}),
\label{eqh0}\\
&&\dot H = -\frac{8\pi G}{2}
(1+\omega_{_M})\rho_{_M},
\label{reh0}
\end{eqnarray}
where the matter-energy density $\rho_{_{_M}}$ and its equation of state 
$\omega_{_M}$
are calculated by using pair 
productions. The cosmological energy density $\rho_{_{_\Lambda}}$ 
governs the spacetime 
inflation rate $H$ and in the meantime produces 
the matter-energy density $\rho_{_{_M}}$, 
whose back reaction, in turn, slows down inflation to its end. 
The obtained results are in agreement with CMB observations. 
Suppose that after reheating the 
matter-energy density is much larger than 
the cosmological energy density, we show due to such back reaction 
that the cosmological term $\rho_{_\Lambda}$ 
tracks down the matter term $\rho_{_M}$ from the reheating 
end to the radiation-matter 
equilibrium, then it varies very slowly, $\rho _{_\Lambda}\propto$ constant, consistently leading to 
the cosmic coincidence in the present time.
The detailed discussions and results of such a scenario  
$\tilde\Lambda$CDM have been presented there.
}

Separating fast components from slow ones  \cite{Chung2019}, we describe 
the slow and fast components' decomposition: scale factor $a=a_{\rm slow}+a_{\rm fast}$, Hubble function $H=H_{\rm slow}+H_{\rm fast}$, cosmological $\tilde\Lambda$ and matter densities $\rho_{_{\Lambda, M, R}}=\rho^{\rm slow}_{_{\Lambda, M, R}}+\rho^{\rm fast}_{_{\Lambda, M, R}}$ and pressures $p_{_{\Lambda, M, R}}=p^{\rm slow}_{_{\Lambda, M, R}}+p^{\rm fast}_{_{\Lambda, M, R}}$. The fast components vary faster in time, but their amplitudes are much smaller than the slow ones. According to the order of small ratio $\lambda$ of fast and slow components, 
the Friedman equations (\ref{friedman}) decompose into two sets. The slow components ${\mathcal O}(\lambda^0)$
obey the same equations as usual Friedman equations (``macroscopic'' ${\mathcal O}(H_{\rm slow}^{-1})$ equations)
\begin{eqnarray}
H_{\rm slow}^2&=&\frac{8\pi G}{3}(\rho_{_M}^{\rm slow}+\rho_{_R}^{\rm slow}+\rho_{_\Lambda}^{\rm slow});\label{sfriedman0}\\
\quad \dot H_{\rm slow}&\approx &-\frac{8\pi G}{2}(\rho_{_M}^{\rm slow} +p_{_M}^{\rm slow}+\rho_{_R}^{\rm slow} +p_{_R}^{\rm slow}),
\label{sfriedman}
\end{eqnarray}
where $H_{\rm slow}=\dot a_{\rm slow}/a\approx 
\dot a_{\rm slow}/a_{\rm slow}$, 
time derivatives $\dot H_{\rm slow}$ and $\dot a_{\rm slow}$ 
relate to the macroscopic 
``slow'' time variation scale ${\mathcal O}(1/H)$. The Equation of the state is $p^{\rm slow}_{_{R,M}}=\omega_{_{R,M}}\rho^{\rm slow}_{_{R,M}}$ for normal radiation and matter (including dark matter) components. They enter the usual dynamics of Universe evolution, i.e., inflation, reheating and standard cosmology.
The faster components ${\mathcal O}(\lambda^1)$ 
obey ``microscopic'' ${\mathcal O}(M^{-1})$ equations \footnote{They differ from the equations obtained by scalar field and potential model in Ref \cite{Chung2019}},
\begin{eqnarray}
H_{\rm fast}&=&\frac{8\pi G}{2\times 3H_{\rm slow}}(\rho_{_M}^{\rm fast}+\rho_{_R}^{\rm fast}+\rho_{_\Lambda}^{\rm fast});\label{ffriedman0}\\ 
\dot H_{\rm fast}&\approx &-\frac{8\pi G}{2}(\rho_{_M}^{\rm fast} +p_{_M}^{\rm fast}+\rho_{_R}^{\rm fast} +p_{_R}^{\rm fast}),
\label{ffriedman}
\end{eqnarray} 
where the fast components of matter density $\rho_{_M}^{\rm fast}$ and pressure $p_{_M}^{\rm fast}$ are due to the non-adiabatic production of massive particle and antiparticle pairs in fast time variation $H_{\rm fast}=\dot a_{\rm fast}/ a_{\rm slow}$ and its time derivative $\dot H_{\rm fast}$. They relate to the microscopic ``fast'' time variation scale ${\mathcal O}(1/M)$. Whereas all slow components approximate as constants ``background'' in ``fast'' time variation. 
The dark-energy equation of state $p_{_\Lambda}=-\rho_{_\Lambda}$ splits into $p_{_\Lambda}^{\rm slow}=\omega_{_\Lambda}\rho_{_\Lambda}^{\rm slow}$ $\mathcal{O}(\lambda^0)$
and $p_{_\Lambda}^{\rm fast}=\omega_{_\Lambda}\rho_{_\Lambda}^{\rm fast}$ $\mathcal{O}(\lambda^1)$, and $\omega_{_\Lambda}$ is at the leading order $\mathcal{O}(\lambda^0)$. Approximation sign ``$\approx$'' in Eqs.~(\ref{sfriedman},\ref{ffriedman}) indicates 
we use $\omega_{_\Lambda}\approx -1$ \footnote{Here, $\omega_{_\Lambda}\approx -1$ is due to time-varying $\tilde\Lambda$ dark energy interacting with matter \cite{Begue2019}. 
In contrast, $\omega_{_\Lambda}=-1$ 
in non-interacting constant $\Lambda$ case.}.

The fast and slow components' separation and coupled Equations (\ref{sfriedman0}-\ref{ffriedman}) are formal and generic. It applies to all Universe's evolution epochs: inflation, reheating and standard cosmology. However, the fast components (\ref{ffriedman0},\ref{ffriedman}) depend on the slow components (\ref{sfriedman0},\ref{sfriedman}) in different evolution epoch. 
In due course, we will discuss what the fast components $\rho_{_M}^{\rm fast}$ and $p_{_M}^{\rm fast}$ in Eqs.~(\ref{ffriedman0}) and (\ref{ffriedman}) are, and how they interact and contribute to the slow components in Friedman equations (\ref{sfriedman0}) and (\ref{sfriedman}).

\comment{Moreover, we adopt the approach \cite{Parker1973} to describe the fast components of matter density $\rho_{_M}^{\rm fast}$ and pressure $p_{_M}^{\rm fast}$ in Eq.~(\ref{ffriedman}). They are due to the non-adiabatic production of massive particle and antiparticle pairs in fast time variation $H_{\rm fast}=\dot a_{\rm fast}/ a_{\rm slow}$ and its time derivative $\dot H_{\rm fast}$.
We find quantum coherent oscillation of fast and microscopic components $H_{\rm fast}$, $\rho_{_\Lambda}^{\rm fast}$, $\rho_{_M}^{\rm fast}$ and $p_{_M}^{\rm fast}$, due to microscopic back reactions at the time scale ${\mathcal O}(M^{-1})$. The quantum pair production and oscillation of $\rho_{_M}^{\rm fast}$ and $p_{_M}^{\rm fast}$ form a macroscopic state of massive pair plasma, contributing to slow macroscopic components $\rho_{_M}^{\rm slow}$ and $p_{_M}^{\rm slow}$ at the time scale ${\mathcal O}(H^{-1})$.
}

\section{Quantum massive pair production and oscillation}\label{qppo}

\subsection{Quantum massive pair production}

In this section, we briefly discuss the Parker and Fulling results \cite{Parker1973} for the gravitational production of a large number of massive particles ($M\gg H_{\rm slow}$) via non-adiabatic processes. We will re-derive the results in the $\tilde\Lambda$CDM (\ref{friedman}) and use them for the fast components of matter density $\rho_{_M}^{\rm fast}$ and pressure $p_{_M}^{\rm fast}$ in Eq.~(\ref{ffriedman0},\ref{ffriedman}). 

In Ref.~\cite{Parker1973}, authors discussed the results for boson fields. It is also valid for fermion fields. A quantised massive scalar matter field inside the Hubble sphere volume 
$V\sim H^{-3}_{\rm slow}$ of Friedman Universe reads
\begin{eqnarray}
\Phi({\bf x},t)&=&\sum_n A_n Y_n({\bf x})\psi_n(t).
\label{qfield}
\end{eqnarray}
Here we consider a massive field $M\gg H_{\rm slow}$ and its modes well localise inside the horizon. The field exponentially vanishes outside the horizon $H^{-1}_{\rm slow}$, i.e., the particle horizon $(a_{\rm slow}H_{\rm slow})^{-1}$ of comoving Hubble radius.
The symbol ``$n$'' 
labels quantum states of physical wave vectors $k_n$, $n=0$ and $k_0=0$ 
for the ground state \footnote{In Ref.~\cite{Parker1973}, the principal quantum number $n$ is the angular momentum number  ``$\ell=0,1,2,\cdot\cdot\cdot$'' and $Y_n({\bf x})=Y_{\ell,m}({\bf x})$ are the four-dimensional spherical harmonics for the closed Robertson-Walker metric and $\Lambda=0$. The ground state is $n=\ell =0$. Here we discuss the case of a flat 
Robertson-Walker metric and $\Lambda\not=0$, for which a massive scalar matter field has no discrete spectra. However, this is not important here since we adopt the Parker-Fulling result (\ref{pairB}) for the ground state $k_0=0$ and $\omega_0=M$, which well localizes inside the horizon. 
}. 
The $A_n$ and $A_n^\dagger$ are time-independent annihilation and creation operators satisfying the commutation relation $[A_n^\dagger,A_n]=\delta_{n,n'}$. 
The time-separate equation for $\psi_n(t)$ is 
\begin{eqnarray}
\partial_t^2\psi_n(t) + \omega_n(t)^2\psi_n(t)=0,\quad 
\omega_n(t)^2=k^2_n+M^2,\label{timeeq}
\end{eqnarray} 
and Wronskian-type condition $\psi_n(t)\partial_t\psi^*_n(t) - \psi^*_n(t)\partial_t\psi_n(t)=i$
in the conformal coupling case. 
Expressing 
\begin{eqnarray}
\psi_n(t)\!&=&\!\frac{1}{(2V\omega_n)^{1/2}}\left(\alpha^*_n(t) e^{-i\int^t\omega_n dt}+\beta^*_n(t) e^{i\int^t\omega_n dt}\right)\label{alphabeta}
\end{eqnarray}
in terms of $\alpha_n(t)$ and $\beta_n(t)$, Equation (\ref{timeeq}) becomes
\begin{eqnarray}
\partial_t\alpha_n(t) &=& C_n e^{-2i\int^t\omega_n dt}\beta_n(t);\nonumber\\
 \partial_t\beta_n(t) &=& C_n e^{2i\int^t\omega_n dt}\alpha_n(t),
\label{eqalphabeta}
\end{eqnarray} 
and $|\alpha_n|^2-|\beta_n|^2=1$, where $C_n\equiv 3H\omega_n^{-2}[k_n^2/3 +M^2/2]$. 
In an adiabatic process for slowly time-varying $H=H_{\rm slow}$, the particle state
$\alpha_n(0)=1$ and $\beta_n(0)=0$ evolves to $|\alpha_n(t)|\gtrsim 1$ and
$|\beta_n(t)|\not=0$. Positive and negative frequency modes get mixed, 
leading to particle productions of probability $|\beta_n(t)|^2\propto e^{-M/H_{\rm slow}}$. 

We will study particle production in non-adiabatic processes of rapidly time-varying 
$H_{\rm fast}$, $\alpha_n$ and $\beta_n$. We focus only on the 
ground state $n=0$ of the lowest-lying massive mode $M\gg H$. 
First, we recall that Parker and Fulling introduced 
transformation \cite{Parker1973}, 
\begin{eqnarray}
A_0=\gamma^*B + \delta B^\dagger,\quad B=\delta A^\dagger_0-\gamma A_0,
\label{bogo}
\end{eqnarray}
$[B,B^\dagger]=1$, and two mixing constants 
obey $|\gamma|^2-|\delta|^2=1$. 
For a given $A_n$ and its Fock space, the state $|{\mathcal N}_{\rm pair}\rangle $ 
is defined by the conditions 
$A_{n\not=0}|{\mathcal N}_{\rm pair}\rangle=0$ and  
\begin{eqnarray}
B^\dagger B |{\mathcal N}_{\rm pair}\rangle ={\mathcal N}_{\rm pair}|{\mathcal N}_{\rm pair}\rangle, \quad {\mathcal N}_{\rm pair}\gg 1.
\label{pairB}
\end{eqnarray}
The $B^\dagger$ and $B$ are time-independent creation and annihilation 
operators of the pair of mixed positive frequency $A_0$ particles and 
negative frequency $A_0^\dagger$ antiparticle. 
The state $|{\mathcal N}_{\rm pair} \rangle $ 
contains ${\mathcal N}_{\rm pair}=1,2,3,\cdot\cdot\cdot$ pairs, 
and it is the ground state 
of non-adiabatic interacting system of fast varying $H_{\rm fast}$ 
and massive pair production and annihilation. 
It is a coherent superposition of states of 
a large occupation number ${\mathcal N}_{\rm pair}$ of particle and anti-particle pairs. In Ref.~\cite{Parker1973}, the authors compared 
it with the BCS condensate state in superconductivity theory and contrasted it with the normal single-particle state. 
In this coherent condensate state $|{\mathcal N}_{\rm pair} \rangle$ and ${\mathcal N}_{\rm pair}\gg 1$, neglecting higher mode 
$n\not=0$ contributions, they obtained the negative quantum pressure 
and positive quantum density of 
coherent pair field, see Eqs.~(59) and (60) of Ref.~\cite{Parker1973},
\begin{eqnarray}
p^{\rm fast}_{_M}&=&-\frac{M(2{\mathcal N}_{\rm pair }+1)}{2\pi^2 V}\Big\{{\rm Re}[\gamma^*\delta(|\alpha|^2+|\beta|^2)]\nonumber\\
&+&(2|\delta|^2+1){\rm Re}(\alpha^*\beta e^{2iMt})\Big\},\label{fastp}\\
\rho^{\rm fast}_{_M}&=&\frac{M(2{\mathcal N}_{\rm pair }+1)}{\pi^2 V}\Big\{{\rm Re}[\gamma\delta^*\alpha\beta)]\nonumber\\
&+&(|\delta|^2+1/2)(|\beta|^2+1/2)\Big\},
\label{fastrho}
\end{eqnarray}
where $\omega_{n=0}=M$, $\alpha_{n=0}=\alpha$ and $\beta_{n=0}=\beta$.
They satisfy the continuity equation of energy-momentum conservation. 
In addition to non-vanishing $|\beta|^2\not=0$, the large occupation number ${\mathcal N}_{\rm pair }\gg 1$ in the coherent state (\ref{pairB}) 
is crucial for the significant gravitational production of massive pairs. It differs from adiabatic particle production in the vacuum state of zero particles.
For a closed Universe case, they adopted the pressure (\ref{fastp}) and density (\ref{fastrho}) for studying the avoidance of cosmic singularity at the beginning of the Universe. 
In their sequent article \cite{Parker1974}, the authors confirm Eqs.~(\ref{fastp}) and 
(\ref{fastrho}) by studying the regularisation of higher mode contributions to the energy-momentum tensor of a massive quantized field 
of closed, flat, and hyperbolic spatial spaces. In the case of heavy particles produced near the Planck scale, the renormalization of high-energy contributions should not be the same as the case of produced light particles in low energies \cite{Xue2019}. The natures of the massive coherent pair state $|{\mathcal N}_{\rm pair}\rangle$ (\ref{pairB}) of the pressure (\ref{fastp}) and density (\ref{fastrho}) are rather generic for non-adiabatic production of massive particles in curved spacetime.
The coherent state $|{\mathcal N}_{\rm pair}\rangle$ (\ref{pairB}) and (\ref{fastp},\ref{fastrho}) should be valid also for $M\gtrsim H$, provided the pair occupation number ${\mathcal N}_{\rm pair}\gg 1$. Note that $p^{\rm fast}_{_M}$ (\ref{fastp}) and $\rho^{\rm fast}_{_M}$ (\ref{fastrho})
represent the quantum pressure and density of massive coherent pair 
state (\ref{pairB}) in short quantum time sales ${\mathcal O}(1/M)$. They do not follow the usual 
equation of the state of classical matter. 

To end this section, we emphasize two points. (i) The 
quantum pressure $p^{\rm fast}_{_M}$ (\ref{fastp}) oscillates, and its value can be positive or negative in oscillations of frequency $1/M$, depending on modes' equation (\ref{eqalphabeta}), superposition coefficients $\gamma,\delta$ (\ref{bogo}) and mass $M$ values. 
The negative value of microscopic time-averaged quantum pressure $p^{\rm fast}_{_M}$ is crucial for forming the coherent condensate state $|{\mathcal N}_{\rm pair}\rangle $ (\ref{pairB}) of a large occupation number ${\mathcal N}_{\rm pair}$ of 
massive particle and anti-particle pairs produced. (ii) Such coherent condensate state $|{\mathcal N}_{\rm pair}\rangle $ occurs only at the ground state $k_0=0$ and $n=0$, i.e., the state of $\ell=0$ spherical $S$-wave. For high angular momentum states $\ell\not=0$, 
the time-averaged pressure becomes non-negative classical values $\propto \ell (\ell +1)$ \cite{Parker1973}. Therefore, the coherent condensate state $|{\mathcal N}_{\rm pair} \rangle $ cannot form for high-energy states $\ell\not=0$. It gives us a lesson that the high-energy modes' renormalization or subtraction prescription for massive particles ($M\sim M_{\rm pl}\gg H$) production is not the same as light particles ($M\ll H \ll M_{\rm pl}$) production.
 
\comment{Therefore higher mode $(k_n\not=0)$ contributions could be neglected. 
Their regularization and corrections will be 
studied in future. In this article, we adopt (\ref{fastp}) and (\ref{fastrho}) as the fast components $\rho^{\rm fast}_{_M}$ and $p^{\rm fast}_{_M}$ in 
Eq.~(\ref{ffriedman}) to find their 
non-adiabatic back-reactions on fast components $H_{\rm fast}$ and 
$\rho^{\rm fast}_{_\Lambda}$.
}

\begin{figure*}[t]
\centering
\begin{center}
\includegraphics[height=5.5cm,width=9.8cm]{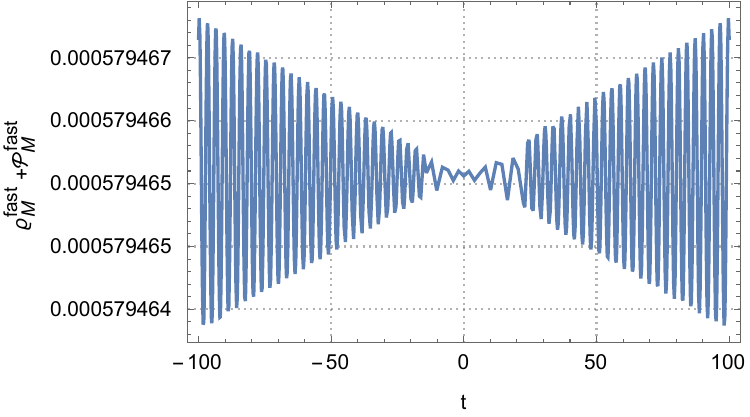}
\caption{We show the quantum pair density and pressure oscillations in microscopic time $t$ in the unit of $M^{-1}$, by using $H_{\rm slow}/M 
\approx 10^{-3}$, $M\simeq 10^{-5}m_{\rm pl}$, 
${\mathcal N}_{\rm pair}\simeq 10^{12}$ and $\delta= 1$, with $C_0 = (3/2)h_{\rm fast}(H_{\rm slow}/M)$ and verified condition $|\alpha|^2-|\beta|^2=1$. 
It shows that a large number of massive pairs 
creates significantly oscillating quantum pressure ${\mathcal P}^{\rm fast}_{_M}$ (\ref{fastp+}) and $\varrho^{\rm fast}_{_M}$ (\ref{fastrho+}) 
in the unit of $\rho_{\rm crit}$, the oscillating amplitudes $\delta\varrho^{\rm fast}_{_M}/\varrho^{\rm fast}_{_M}$ and $\delta {\mathcal P}^{\rm fast}_{_M}/{\mathcal P}^{\rm fast}_{_M}$ are about ${\mathcal O}(10^{-3})$. For a long time, 
the coherent oscillations approach stable configurations in time. 
For more details and figures, for instance, the fast components $\varrho^{\rm fast}_{_\Lambda}$, $h_{\rm fast}$ and $\dot h_{\rm fast}$, see Fig.~\ref{reh-detailosci1+} in Appendix. Note that the pair number ${\mathcal N}_{\rm pair}$, mass scales $M$ and $H_{\rm slow}$ values differ from those used for inflation see Fig.~1 in Ref.~\cite{Xue2021}.}
\label{reh-osci+}
\end{center}
\vspace{-2em}
\end{figure*}

\subsection{Quantum massive pair oscillation}

Following their approach for the ground state $k_n=0$, we arrive at the same quantum pressure (\ref{fastp}) and density (\ref{fastrho}) 
in the $\Lambda$CDM. 
In our case, we consider the state (\ref{pairB}) 
as a coherent condensate state of very massive $M\gg H_{\rm slow}$ and 
large number ${\mathcal N}_{\rm pair}\gg 1$ pairs. Therefore, $M(2{\mathcal N}_{\rm pair}+1)$ in Eqs.~(\ref{fastp},\ref{fastrho}) 
can be larger than the Planck mass, and higher mode $(k_n\not=0, n\not=0)$ contributions can be neglected. Their regularisation and corrections will be 
studied in future.  
In this article, we adopt $p^{\rm fast}_{_M}$ (\ref{fastp}) and $\rho^{\rm fast}_{_M}$ and (\ref{fastrho}) as the fast components in 
Eqs.~(\ref{ffriedman0},\ref{ffriedman}) to find 
their non-adiabatic back-reactions on 
fast components $H_{\rm fast}$ and 
$\rho^{\rm fast}_{_\Lambda}$.

We will study the reheating epoch when the Hubble scale and pair mass are very much smaller than the Planck scale, i.e., $H_{\rm slow}<M \ll m_{\rm pl}$ and ${\mathcal N}_{\rm pair}\gg 1$. 
Therefore, in the unit of the mass $M$ and the critical density $\rho_{\rm crit}=3m_{\rm pl}^2H^2_{\rm slow}$, we express the dimensionless quantum pressure (\ref{fastp}) and density (\ref{fastrho}) as  \footnote{In the previous article \cite{Xue2021}, we use the reduce Planck
mass $m_{\rm pl}$ and energy density $m^4_{\rm pl}$
as the unit for studying inflation.}
\begin{eqnarray}
{\mathcal P}^{\rm fast}_{_M}&=&-\frac{\bar M H_{\rm slow}}{6\pi^2 m_{\rm pl}}\Big\{{\rm Re}[\gamma^*\delta(|\alpha|^2+|\beta|^2)]
+(2|\delta|^2+1){\rm Re}(\alpha^*\beta e^{2iMt})\Big\},\label{fastp+}\\
{\varrho}^{\rm fast}_{_M}&=&+\frac{\bar M H_{\rm slow}}{3\pi^2 m_{\rm pl}}\Big\{{\rm Re}[\gamma\delta^*\alpha\beta)]+(|\delta|^2+1/2)(|\beta|^2+1/2)\Big\},
\label{fastrho+}
\end{eqnarray}
where $\bar M\equiv (2{\mathcal N}_{\rm pair }+1)(M/m_{\rm pl})$.
The fast component equations (\ref{ffriedman0},\ref{ffriedman}) become,
\begin{eqnarray}
h_{\rm fast}&=&\frac{1}{2}(\varrho_{_M}^{\rm fast}
+\varrho_{_\Lambda}^{\rm fast});\nonumber\\ 
\dot h_{\rm fast}&=&-\frac{3}{2}(\varrho_{_M}^{\rm fast} 
+{\mathcal P}_{_M}^{\rm fast}),
\label{ffriedman+}
\end{eqnarray}
where $h_{\rm fast}\equiv H_{\rm fast}/H_{\rm slow}$ and $\varrho_{_\Lambda}^{\rm fast}\equiv \rho_{_\Lambda}^{\rm fast}/\rho_{\rm crit}$. Here we only consider the fast components of massive particle productions and oscillations inside the Horizon and neglect the fast components of light particles.

Using negative ${\mathcal P}^{\rm fast}_{_M}$ (\ref{fastp+}) and positive definite $\varrho^{\rm fast}_{_M}$ (\ref{fastrho+}), we search for a solution of fast component equation (\ref{ffriedman+}) and quantum fluctuating mode equations (\ref{eqalphabeta}) in the period $[-t,t]$ of the microscopic time $t\sim H^{-1}_{\rm fast}$. The period is around the macroscopic time $t_{\rm slow}\sim H^{-1}_{\rm slow}$, when the slow components $a_{\rm slow}$, $H_{\rm slow}$, $\rho^{\rm slow}_{_{M,\Lambda}}$ and $p^{\rm slow}_{_{M,\Lambda}}$ are determined by 
the Friedman equations (\ref{sfriedman0},\ref{sfriedman}). 
The integrals 
$\int^t\omega_n dt$ are over the microscopic time $t$  characterised by 
the time scale $1/M$. Its lower limit is $t=0$ by setting 
$t_{\rm slow}=0$ as a reference time, when $a_{\rm fast}(0)=0$,
\begin{eqnarray}
H_{\rm fast}(0)=\dot a_{\rm fast}/a_{\rm slow}=0;\quad \alpha(0)=1,\quad \beta(0)=0.
\label{initial}
\end{eqnarray}
The real value $\gamma^*\delta$ condition in Eqs.~(\ref{fastp+}),\ref{fastrho+}) leads to the time symmetry:
$a^{\rm fast}(t)=a^{\rm fast}(-t)$, $\alpha(t)=\alpha^*(-t)$ 
and $\beta(t)=\beta^*(-t)$ \cite{Parker1973}. When $t\leftrightarrow -t$, positive and negative frequency modes interchange. 
\comment{In Ref.~\cite{Parker1973}, $a_{\rm slow}=0$, $H_{\rm slow}=0$ (i.e.,
$H=H_{\rm fast}$, $a=a_{\rm fast}$) and a small spherical volume $V\sim (H_{\rm fast})^3$ at the cosmic origin was adopted for studying the avoidance of cosmic singularity for the $\rho_{_\Lambda}=0$ and curved Universe.} 
Here we use $a_{\rm slow}\not=0$, $H_{\rm slow}\not=0$ and co-moving radius $(Ha)^{-1}\approx (H_{\rm slow}a_{\rm slow})^{-1}$ of Hubble volume $V\sim H_{\rm slow}^{-3}$.

In microscopic time $t$ of unit $M^{-1}$, we numerically solve 
non-linearly coupled Eqs.~(\ref{eqalphabeta}) and (\ref{fastp+}-\ref{ffriedman+}) with the initial condition (\ref{initial}). We report the results in Fig.~\ref{reh-osci+} and
details in Fig.~\ref{reh-detailosci1+} 
of Appendix. Similar to the previous results \cite{Xue2021}, we find that in the quantum period of microscopic time $t$, the negative quantum pressure ${\mathcal P}^{\rm fast}_{_M} < 0$ and back-reaction effects lead to the {\it quantum pair oscillation} in a time characterised by the frequency $\omega\sim M$. The small $a_{\rm fast}(t)$ varies around $a_{\rm slow}$ at $t_{\rm slow}\equiv 0$.  
The massive pairs' density and pressure ($\varrho^{\rm fast}_{_M},{\mathcal P}^{\rm fast}_{_M}$) oscillate coherently with the spacetime fields $(h_{\rm fast},\dot h_{\rm fast},\varrho^{\rm fast}_{_\Lambda})$ 
oscillations. Their oscillatory structures imply
a quantum back-and-forth process in microscopic time scale ${\mathcal O}(1/M)$
\begin{eqnarray}
{\mathcal S}\Leftrightarrow  \bar FF
\label{bfrhoq}
\end{eqnarray}
between spacetime fields ${\mathcal S}(h_{\rm fast},\dot h_{\rm fast},\varrho^{\rm fast}_{_\Lambda})$ and massive particle pairs $\bar FF(\varrho^{\rm fast}_{_M},{\mathcal P}^{\rm fast}_{_M})$, previously discussed \cite{Xue2019}.
These results show the highly non-adiabatic and complex nature of massive pair-production processes and collective oscillations. Attributed to complex back-and-forth reactions at the scale ${\mathcal O}(1/M)$, the quantum massive pair oscillation (\ref{bfrhoq}) cannot be described by oscillating scalar fields with polynomial potential.

As shown in Figs.~\ref{reh-osci+} and \ref{reh-detailosci1+}, for the microscopic time $t\gg 1/M$, the positive quantum pair density $\varrho^{\rm fast}_{_M}> 0$ indicates particle creations without $e^{-M/H}$ suppression. 
It is consistent with increasing Bogoliubov 
coefficient $|\beta(t)|^2$ that mixes positive and negative energy modes. 
Observe that $\varrho^{\rm fast}_{_M}> |{\mathcal P}^{\rm fast}_{_M}|$ and the sum $\varrho^{\rm fast}_{_M}+ {\mathcal P}^{\rm fast}_{_M} >0$ is positive definite, leading to the decreasing $h_{\rm fast}(t)$ (\ref{ffriedman+}). As a consequence,   
for positive time ($t>0$) increasing (forward the future), the fast components $h_{\rm fast}$ 
and $\varrho^{\rm fast}_{_\Lambda}$ decrease, in order for pair production. Whereas for negative time ($t<0$) increasing (backward the past), $h_{\rm fast}$ and $\varrho^{\rm fast}_{_\Lambda}$ increases, due to pair annihilation. 
In both situations, the $\rho^{\rm fast}_{_\Lambda}$ is negative and $\rho^{\rm fast}_{_M}$ is positive, as required by the energy conservation (\ref{ffriedman+}) in the massive pairs' production via fast oscillating $h_{\rm fast}$ and $\dot h_{\rm fast}$. Equations $\rho_{_{\Lambda, M}}=\rho^{\rm slow}_{_{\Lambda, M}}+\rho^{\rm fast}_{_{\Lambda, M}}$ imply the dark energy $\rho_{_\Lambda}$ decreases and matter $\rho_{_M}$ increases, where the slow components $\rho^{\rm slow}_{_{\Lambda, M}}$ are fixed values at $t=t_{\rm slow}\equiv 0$. In this sense, dark energy converts to massive pairs (matter) in a microscopic time scale, whereas the case for a macroscopic time scale will be studied in Sec.~\ref{mppa}.
It is our finding that the energy density and pressure (\ref{fastp+},\ref{fastrho+}) of the condensate state $|{\mathcal N}_{\rm pair}\rangle$ coherently interacts and exchanges energy with the fast components of spacetime variation (\ref{ffriedman+}). Such 
the back-and-forth reaction at the 
scale $1/M$ was not studied in Ref.~\cite{Parker1973} for the $|{\mathcal N}_{\rm pair}\rangle$ condensate state's energy density (\ref{fastrho}) and pressure (\ref{fastp}). In our solutions, we find the quantum pressure ${\mathcal P}^{\rm fast}_{_M}$ and its time average are negative see Fig.~\ref{reh-detailosci1+} in Appendix. It implies the possibility that the condensate state $|{\mathcal N}_{\rm pair} \rangle$ retains while it coherently interacts with the fast components of spacetime horizon variation
(\ref{bfrhoq}).

\comment{On the other classic viewpoints, note that the kinetic energy density (\ref{fastrho+}) of particle and antiparticle pairs 
is positive. The negative pressure (\ref{fastp+}) implies an effectively negative potential (attractive force) between particles and antiparticles. It is why the quantum pair oscillation proceeds along with oscillating spacetime fields $(h_{\rm fast},\dot h_{\rm fast},\varrho_{_\Lambda}^{\rm fast})$ in Eqs.~(\ref{ffriedman+}).}
This phenomenon is dynamically analogous 
to the plasma oscillation of electron-positron pair production in 
an external alternating electric field $E$ \cite{Kluger1991}. The pair production rate 
is not exponentially suppressed by $e^{-\pi M^2/E}$ \cite{Ruffini2010}. 
The coherent plasma state of electron-positron pairs is analogous to the 
coherent pair state $|{\mathcal N}_{\rm pair}\rangle $ (\ref{pairB}) and quantum pair oscillation, shown in Fig.~\ref{reh-osci+}.

Such massive and semi-classical state of large occupation number (\ref{pairB}) and quantum pair oscillation (Fig.~\ref{reh-osci+}) well localises inside the horizon.  
They persist throughout the entire Universe's history, independent of slow components $H_{\rm slow}$ and $\rho^{\rm slow}_{_{\Lambda, M, R}}$ values in Friedman equations (\ref{sfriedman0},\ref{sfriedman}).
However, the pair mass $M$ (oscillating frequency $\omega\approx M$) and pair number ${\mathcal N}_{\rm pair}$ depend on slow components' values in the Friedman equations (\ref{sfriedman0},\ref{sfriedman}). It is necessary and deserves to proceed with further studies.

\comment{This analogy motivates us to model the quantum coherent pair state  
$|{\mathcal N}_{\rm pair}\rangle $ (\ref{pairB}) and oscillating dynamics (Fig.~\ref{reh-osci+}) as a quasi-classical plasma state
of effective energy density and pressure to study its impacts on the Friedman equation (\ref{sfriedman}) in the $\tilde\Lambda$CDM scenario. 
}

\section{Massive pair plasma state and holographic hypothesis}\label{mpp}

\subsection{Reasons for a macroscopic description}

We see the non-adiabatic and back-reacting phenomena of massive pairs' production and oscillation at a time scale ${\mathcal O}(1/M)$. 
How the fast oscillating components ${\mathcal P}^{\rm fast}_{_M}$ (\ref{fastp+}) and $\varrho^{\rm fast}_{_M}$ (\ref{fastrho+}) 
in Eqs.~(\ref{ffriedman0},\ref{ffriedman}) couple to the slow components in Friedman equations (\ref{sfriedman0},\ref{sfriedman}). It is a difficult task to simultaneously analyze ${\mathcal O}(1/M)$ and ${\mathcal O}(1/H)$ back-reaction dynamics even numerically since two scales $M\gg H$ are very different. To deal with this difficulty, we adopt two approximate steps. First, due to 
nontrivial time-averaged values of fast components over the microscopic time $t\gg 1/M$, we assume the massive pair production and oscillation form a {\it massive pair plasma state} in the macroscopic time scale. 
We model such a semi-classical state as a perfect fluid by defining the effective density and pressure. Second, we discuss how it back-and-forth interacts and contributes to the slow components in the Friedman equations.

Figure \ref{reh-osci+} shows that massive pair quantum pressure ${\mathcal P}^{\rm fast}_{_M}$ (\ref{fastp+}) and density $\varrho^{\rm fast}_{_M}$ (\ref{fastp+}) rapidly oscillate with the fast components $h_{\rm fast}$ and $\varrho^{\rm fast}_{_\Lambda}$ (\ref{ffriedman+}) in microscopic time. Their oscillating amplitudes are significantly large and not dampening in time. It, therefore, expects to form 
{\it a massive pair plasma state} in a macroscopic time and space. 
However, to study its effective impacts on the classical Friedman equations (\ref{sfriedman}), we have to discuss two problems stemming from the scale difference 
$M\gg H_{\rm slow}$. 
\begin{enumerate}[(i)]
\item 
First, the different time scales. It is impossible to even numerically integrate slow and fast component coupled equations (\ref{sfriedman},\ref{ffriedman}) due to their vastly different time scales. On this aspect, we consider the
fast-component averages $\langle\cdot\cdot\cdot\rangle$ over the microscopic time scale. Figure \ref{reh-osci+} shows $\varrho^{\rm fast}_{_M} + {\mathcal P}^{\rm fast}_{_M} >0$, which does not oscillate alternatively between negative and positive values. 
Its time average $\langle\varrho^{\rm fast}_{_M}+ {\mathcal P}^{\rm fast}_{_M}\rangle$ does not vanish. Other 
fast-component averages do not vanish as well. 
Fast-component averages have another time scale $\tau_{_M}$ (\ref{prate}) in response to the slow horizon variations $H_{\rm slow}$ in macroscopic time. It is a kind of ``relaxation'' time scale and differs from the oscillating one $1/M$. Therefore, in principle, fast-component averages possibly affect the Friedman equation at the macroscopic time scale. In practice, 
the appropriate modelling of fast-component averages can avoid the difficulty 
of vastly different scale dynamics in calculations, and the scenario becomes tractable. However, we have to check its self-consistency with observation. 
\item Second, the spatial distribution. We do not know the 
spatial distribution of the massive pair condensate state $|{\mathcal N}_{\rm pair}\rangle$. Namely, we do not know the radial dependence of the quantum pressure and density (\ref{fastp},\ref{fastrho}) or (\ref{fastp+},\ref{fastrho+}), since the Ref.~\cite{Parker1973} authors obtained them 
by using the vacuum expectation value of field $\Phi({\bf x},t)$ (\ref{qfield}) energy-momentum tensor integrated over the entire space. There, they studied the cosmic singularity problem in the Universe beginning $M\gtrsim H$ case, namely the massive mode wavelength $M^{-1}$ is comparable with the horizon size $H^{-1}$. Here, we study the case $M\gg H\approx H_{\rm slow}$, namely the massive mode wavelength $M^{-1}$ is much smaller than the horizon size $H^{-1}_{\rm slow}$. As a hypothesis, we speculate that the massive pair condensate state $|{\mathcal N}_{\rm pair}\rangle$ and the coherent oscillation (\ref{bfrhoq}) with $H_{\rm fast}$ and $\dot H_{\rm fast}$ spatially 
localize nearby the horizon following the holographic principle \cite{Hooft1993, Susskind1995, Cohen1999}. 
The arguments are the following. (a) Such condensate state $|{\mathcal N}_{\rm pair}\rangle$ and oscillation (\ref{bfrhoq}) collectively couple with the fast components $H_{\rm fast}$ and $\dot H_{\rm fast}$, which are quantum modes at short wavelengths ($1/M$). These modes should associate with the horizon surface according to the holographic principle. (b) The massive pair condensate state $|{\mathcal N}_{\rm pair}\rangle$ is a ground state of a spherically symmetric $S$ wave. (c) Such a very massive state of the radial size about $M^{-1}\ll H^{-1}_{\rm slow}$ is inside the horizon $H^{-1}_{\rm slow}$ of the Friedman Universe, whose isotropic homogeneity extends up to the horizon.
\end{enumerate} 
Based on these hypotheses, we will introduce a holographic and massive pair plasma state that gives an effective description of the condensate state $|{\mathcal N}_{\rm pair}\rangle$ and coherent oscillation (\ref{bfrhoq}) at macroscopic space and time scales. 

\subsection{Effective description of a massive pair plasma state}

Based on these considerations, we assume a massive pair plasma state forms in a macroscopic time scale. We describe such macroscopic state as a perfect fluid state of effective number $n^H_{_M}$ and energy $\rho^H_{_M}$ densities \footnote{Here we present the simplest state, and it can be a more complex state of massive pair plasma.
}
\begin{eqnarray}
\rho^H_{_M} \equiv  2\chi  m^2 H^2_{\rm slow},\quad n^H_{_M} \equiv   \chi  m H^2_{\rm slow};\quad m^2 \equiv 
\sum_fg_d^fM^2_f,
\label{apdenm}
\end{eqnarray}
and pressure $p^H_{_M}=\omega^H_{_M}
\rho^H_{_M}$. The $\omega^H_{_M}\approx 0$ for $m\gg H_{\rm slow}$
and its upper limit is $1/3$. The introduced mass parameter $m$ represents possible particle masses $M_f$, degeneracies $g_d^f$ 
and the mixing coefficient $\delta$ (\ref{bogo}). 
The degeneracies $g_d^f$ plays the same role of pair number ${\mathcal N}_{\rm pair}$ in 
Eqs.~(\ref{fastp},\ref{fastrho}) or 
(\ref{fastp+},\ref{fastrho+}). 
The pair masses $M_f$ are smaller than the Planck mass $M_{\rm pl}$, but the mass parameter $m$ can be larger than $M_{\rm pl}$ for a large occupation number ${\mathcal N}_{\rm pair}\gg 1$ or degeneracy $g_d^f\gg 1$. Note that the massive pair plasma state contains (i) unstable massive pairs that couple and decay to light particles; (ii) stable massive pairs with gravitational interaction only. Besides, one should differ the massive pair plasma state density $\rho^H_{_M}$ (\ref{apdenm}) from the normal matter or radiation (including dark matter) density $\rho_{_{M,R}} \propto (1/a)^{3(1+\omega_{_{M,R}})}$. 
The reason is that the massive pair plasma state (\ref{apdenm}) attributes to quantum 
pair production and oscillation, which couple to the oscillating spacetime fields ${\mathcal S}(h_{\rm fast},\dot h_{\rm fast},\varrho^{\rm fast}_{_\Lambda})$ of 
Hubble function and dark energy. To some extent, we may consider the massive pair plasma state as an ``equilibrium state'' between quantum massive pairs and spacetime field oscillations.
  
Following the previous subsection discussions, we explain 
the reasons why the densities (\ref{apdenm}) 
are proportional to $\chi m H^2_{\rm slow}$, rather than $H^3_{\rm slow}$ 
of the entire Hubble volume $V$. 
The ``surface area'' factor $H^2_{\rm slow}$ 
is attributed to the 
spherical symmetry of Hubble volume. The ``radial size'' factor $\chi m$ is
the layer width $\lambda_m $ introduced as an effective parameter to
describe the properties: (i) for $m\gg H_{\rm slow}$ the massive pair plasma state
localizes as a spherical layer near to the horizon; (ii) the layer radial width $\lambda_m < H^{-1}_{\rm slow}$
depends on the massive pair plasma oscillation dynamics \footnote{It may also include self-gravitating dynamics due to the pair plasma state being very massive.}, 
rather than the $H_{\rm slow}$ dynamics govern by the Friedman equations (\ref{sfriedman}). 
The width parameter $\chi$ expresses the layer width $\lambda_m = (\chi m)^{-1} \gg 1/m$,
\begin{eqnarray}
\lambda_m=(\chi m)^{-1} < H^{-1}_{\rm slow},\quad  1\gg \chi > (H_{\rm slow}/m).
\label{chi}
\end{eqnarray}
Note that studying the prescription of high-energy modes' subtraction for $m\gg H$, 
we approximately obtained the mean density $n^H_{_M} \approx \chi m H^2$ (\ref{apdenm}) and $\chi\approx 1.85\times 10^{-3}$ by studying massive fermion pair productions in a De Sitter spacetime of constant $H$ and scaling factor $a(t)=e^{iHt}$ \cite{Xue2019, Xue2020}. We adopt this $\chi$ value for numerical calculations in the present article. 

Since the parameters $m$ and $\chi m$ represent time-averaged values over 
fast time oscillations of massive pair plasma state, we consider $m$ and 
$\chi m$ as approximate constants 
in slowly varying macroscopic time for the Friedman equations. However, the typical $m$ and $\chi m$ values should be 
different for Universe evolution epochs since the fast-component 
equations for massive pair productions and oscillations depend on 
the $H_{\rm slow}$ value, see Sec.~\ref{qppo}.
We used the parameter $m^*$ for inflation, $\hat m$ for reheating and $m_{_M}$ for the epochs 
after reheating. We will fix these parameter values by observations.  

To end this section, we have to point out that (i) the pressure $p^H_{_M}$ and density $\rho^H_{_M}$ (\ref{apdenm}) are effective descriptions of the massive pair plasma state in macroscopic scales, that result from the coherence condensation state (\ref{pairB},\ref{fastp},\ref{fastrho}) and oscillating dynamics (Fig.~\ref{reh-osci+}) in microscopic scales; (ii) they contribute to the ``slow'' components $\rho^{\rm slow}_{_M}$ 
and $p^{\rm slow}_{_M}$ in the ``macroscopic'' ${\mathcal O}(H^{-1}_{\rm slow})$
Friedman equations (\ref{sfriedman0},\ref{sfriedman}). It means that in the Friedman equations (\ref{sfriedman0},\ref{sfriedman}), the matter density and pressure terms $\rho^{\rm slow}_{_M}$ and $p^{\rm slow}_{_M}$ contain (a) the normal matter state contributions 
and (b) the massive pair plasma state contributions.
This will be clarified in the next Section.
We shall study the massive pair plasma state effects on each epoch of the Universe's evolution. Here we investigate its impact on reheating. 

After we adopt the effective description of massive pair plasma state (\ref{apdenm}), 
the quantum massive pair oscillation  (Fig.~\ref{reh-osci+}) 
of fast components $(\varrho^{\rm fast}_{_M},{\mathcal P}^{\rm fast}_{_M}, h_{\rm fast},
\dot h_{\rm fast},\varrho^{\rm fast}_{_\Lambda})$ and $p^{\rm fast}_{_\Lambda}\approx - \rho^{\rm fast}_{_\Lambda}$ details at the scale ${\mathcal  O}(1/M)$
average out and become irrelevant for the macroscopic scale ${\mathcal  O}(\tau_{_H})$ and ${\mathcal  O}(\tau_{_M})$ processes: inflation, reheating and standard cosmology. 
The relevant quantities and equations are massive pair plasma state $p^H_{_M}=\omega^H_{_M}\rho^H_{_M}$ (\ref{apdenm}) and slow components obeying Friedman equations (\ref{sfriedman0},\ref{sfriedman}), 
and their interacting equation (\ref{rateeqd}). The final results depend only on the plasma state (\ref{apdenm}) with the mass $m$ and width $\chi$ parameters. 
Henceforth we ignore the ``fast" components, sub-script, and super-scripts ``slow'' will be dropped.    

\section{Back-and-forth process and cosmic rate equation}\label{mppa}

In Sec.~\ref{qppo}, we show massive pairs' production and annihilation at time scale ${\mathcal O}(1/M)$ via quantum pair oscillations, and dark energy effectively converts to massive pairs for the microscopic time $t\gg 1/M$. These are the quantum back-and-forth process (\ref{bfrhoq}). By non-vanishing averages over microscopic time, these microscopic back-reaction processes should impact the classical and slow components in Friedman's equations. 
Using massive pair plasma state (\ref{apdenm}), we will discuss how to effectively describe the back-and-forth process between massive pairs and spacetime fields $(H,\rho_{_\Lambda})$ at a macroscopic time scale $(1/H)$.

\subsection{Stable massive pairs and cosmic rate equation}\label{mppas}

We discuss here how
the massive pair plasma state (\ref{apdenm}) back-reacts and contributes to the slow components in Friedman equations. 
First, we introduce the mean pair production rate $\Gamma_M$ to describe the massive pair plasma state variation
as the macroscopic time $t$ varies. 
We estimate the total number of particles produced inside the Hubble sphere $N\approx n^H_{_M}H^{-3}/2$ and mean pair production rate w.r.t.~macroscopic time variation $dt$,
\begin{eqnarray}
\Gamma_M &=& \frac{dN}{2\pi dt}\approx \frac{\chi m}{4\pi} \epsilon, \quad \tau^{-1}_{_M}=\Gamma_M. \label{prate} 
\end{eqnarray}
It is 
in terms of the parameter 
$\chi m$ (\ref{chi}) and Universe evolution $\epsilon$-rate defined as,
\begin{eqnarray}
\epsilon &\equiv& -\frac{\dot H}{H^2}=\frac{3}{2}\frac{(1+\omega_{_M})\rho_{_M}+ (1+\omega_{_R})\rho_{_R}}{\rho_{_\Lambda}+\rho_{_M}+\rho_{_R}}.
\label{dde0}
\end{eqnarray}
The second equation comes from the Friedman equations (\ref{sfriedman0},\ref{sfriedman}).
The asymptotic values $\epsilon \approx  0$, $\epsilon \approx  2$ and $\epsilon \approx  3/2$ correspond to the dark-energy (inflation), radiation, and matter dominant epochs, respectively.

The massive pair plasma state $\rho^H_{_M}$ (\ref{apdenm}) effectively represents an equilibrium state of quantum pair and spacetime field oscillations (\ref{bfrhoq}). It
not only depends on the Hubble function $H$, but also contributes to the normal matter density $\rho_{_M}$.
Back reactions must act on $\rho^H_{_M}$, 
when $H$ and $\rho_{_M}$ vary in time following the Friedman equations. Moreover, the massive pair plasma state variation time scale $\tau_{_M}=\Gamma^{-1}_{M}$ is 
smaller than the normal matter density \footnote{For the sake of brief notation, $\rho_{_{M}}$ stands for $\rho_{_{M,R}}$ in this section} $\rho_{_{M}}$ variation time scale $\tau_{_H}=1/H$. The difference $\tau_{_H}>\tau_{_M}$
implies the back-and-forth interaction between the massive pair plasma state density and 
the normal matter density
\begin{eqnarray}
\rho^H_{_M}\Leftrightarrow \rho_{_M}, 
\label{bfrho}
\end{eqnarray}
during the Universe's evolution. The process is induced by quantum pair oscillation coherently with fast oscillating components of the Hubble function and dark energy. 

To model such dynamics (\ref{bfrho}), we recall the rate equation for the back-and-forth process $e^+e^-\Leftrightarrow\gamma\gamma$ \cite{Kolb1990, Lee1977, Ruffini1999b, Ruffini2000}:
\begin{eqnarray}
\frac{dn_{e^+e^-}(t)}{dt} +3 H n_{e^+e^-}(t) = \langle\sigma v \rangle \Big(n^2_{e^+e^-}\big|_{\rm eq}-n^2_{e^+e^-}\Big),
\label{rateee}
\end{eqnarray}
where $n_{e^+e^-}(t)$ is the electron and positron pair density governed by the macroscopic time scale $H^{-1}$ evolution. While $n_{e^+e^-}\big|_{\rm eq}$ is the density of electrons and positrons in equilibrium with two photons $n_{\gamma\gamma}\big|_{\rm eq}$ in microscopic time scale $(\langle\sigma v \rangle n_{e^+e^-})^{-1}$, namely $n_{e^+e^-}\big|_{\rm eq}\approx n_{\gamma\gamma}\big|_{\rm eq}$. The RHS represents the averaged interacting rate $dN/dt\approx \langle\sigma v \rangle n_{e^+e^-}$ for microscopic detail balance between $n_{e^+e^-}(t)$ and $n_{e^+e^-}\big|_{\rm eq}$. They are coupled for $n_{e^+e^-}\big|_{\rm eq}\approx n_{e^+e^-}$ and decoupled for $n_{e^+e^-}\big|_{\rm eq}\ll n_{e^+e^-}$.

We make the following analogies: $n_{e^+e^-}\leftrightarrow \rho_{_M}$, $n_{e^+e^-}\big|_{\rm eq}\leftrightarrow \rho^H_{_M}$ and photons $n_{\gamma\gamma}\big|_{\rm eq}$ correspond to fast oscillating components of the Hubble function and dark energy. This analogy motivates us to propose an effective cosmic rate equation 
\begin{eqnarray}
\dot\rho_{_M}+ 3(1+\omega_{_M}) H\rho_{_M} &=& \Gamma_M(\rho_{_M}^H - \rho_{_M}) - \Gamma_M^{^{\rm de}}\rho_{_M},
\label{rateeqd}
\end{eqnarray}
 of the Boltzmann type for the the back-and-forth  $\rho_{_M}$ and $\rho^H_{_M}$ interaction (\ref{bfrho}) 
in the Universe's evolution. It represents a general conservation law of dark energy and matter, including massive pair plasma state $\rho_{_M}^H$ (\ref{apdenm}) with the production rate (\ref{prate}). 
The term $3(1+\omega_{_M}) H\rho_{_M}$ of the time scale $[3(1+\omega_{_M}) H]^{-1}$ represents the space-time expanding effect on the density $\rho_{_M}$. While $\Gamma_M \rho_{_M}^H$ is the source term and 
$\Gamma_M\rho_{_M}$ is the depletion term. 
The detailed balance term 
$\Gamma_M(\rho_{_M}^H - \rho_{_M})$ indicates how two densities $\rho_{_M}^H$ and $\rho_{_M}$ of different time scales couple together. The ratio $\Gamma_M/H> 1$ indicates the coupled case, and  $\Gamma_M/H < 1$ indicates the decoupled case. 
The last term $\Gamma^{^{\rm de}}_M\rho_{_M}$ represents unstable massive pairs' decay to relativistic particle pairs $\bar\ell\ell$, and 
the decay rate and time are given by
\begin{eqnarray}
\Gamma^{^{\rm de}}_M = g^2_{_Y} m,\quad \tau_{_R} =(\Gamma^{^{\rm de}}_M)^{-1}, \quad (\bar FF\Rightarrow \bar\ell\ell)
\label{Mdecayr} 
\end{eqnarray}
where $g_{_Y}$ is the Yukawa coupling between the massive pairs ($\bar FF$) and
relativistic particles. It is important to note that the decay rate 
$\Gamma^{^{\rm de}}_M$ (\ref{Mdecayr}) depends not only on the Yukawa coupling $g_{_Y}$ but also on the phase space of final states. While for stable massive pairs, the decay rate $\Gamma^{^{\rm de}}_M$ is zero.

The combination of cosmic rate equation (\ref{rateeqd}) and Friedman equations (\ref{sfriedman}) yields
\begin{eqnarray}
\dot\rho_{_\Lambda}&=& -  \Gamma_M \left(\rho^H_{_M} - \rho_{_M}\right) + \Gamma_M^{^{\rm de}}\rho_{_M}= -\delta Q,
\label{rhomm}\\
\dot\rho_{_M} &+& 3(1+\omega_{_M}) H\rho_{_M}=\delta Q
\label{rholm}
\end{eqnarray}
where $\delta Q\equiv \Gamma_M \left(\rho^H_{_M} - \rho_{_M}\right) -\Gamma_M^{^{\rm de}}\rho_{_M}$, representing the interaction and exchange between dark energy and normal matter via the massive pair plasma state $\rho^H_{_M}$.
To discuss this in some more detail, we ignore 
the decay term $\Gamma_M^{^{\rm de}}\rho_{_M}$. It is negligible for unstable pairs, provided $\Gamma_M\gg\Gamma_M^{^{\rm de}}$. We point out four particular cases: 
\begin{enumerate}[(i)]
\item  Recall the results \cite{Xue2021} for the inflation epoch, when the radio $\Gamma_M/H\propto\epsilon_* \ll 1$, $\rho_{_\Lambda}\gg \rho^H_{_M}$ and $\rho^H_{_M}\gg \rho_{_M}$ \footnote{In Ref.~\cite{Xue2021}, we approximately neglect $\rho_{_M}$ and cosmic rate equation (\ref{rateeqd}) to obtain an analytical solution.}.
Equations (\ref{rhomm},\ref{rholm}) become $\dot\rho_{_\Lambda}\approx - \Gamma_M \rho^H_{_M}\lesssim 0$ showing
$\rho^H_{_M}$ production costs dark energy, but it adds into matter-energy $\dot\rho_{_M} + 3(1+\omega_{_M}) H\rho_{_M}\approx \Gamma_M \rho^H_{_M}\gtrsim 0$. 
The exchange rate $\delta Q\approx \Gamma_M \rho^H_{_M}\gtrsim 0$ is positive and small. It yields $\dot\rho_{_\Lambda}\lesssim 0$ and $\dot H\lesssim 0$, i.e., slow-rolling dynamics for inflation. Dark energy converts slowly
to matter till inflation ends when $\Gamma_M/H\approx 1$.
\item In a short pre-reheating episode, when $\Gamma_M /H \gg 1$ and $\rho_{_\Lambda}>\rho^H_{_M}>\rho_{_M}$, the exchange rate $\delta Q\gg 1$ is large. Dark energy rapidly converts to matter till $\rho^H_{_M}\approx \rho_{_M}>\rho_{_\Lambda}$. The conversion is very efficient. Details will be in Sec.~\ref{pre}.
\item The coupled case is $\Gamma_M /H > 1$ and $\rho^H_{_M}\approx \rho_{_M}$, when $\rho^H_{_M}$ 
tightly couples with $\rho_{_M}$ in the Universe evolution of the Hubble time scale $\tau_{_H}$. Dark energy and matter exchange rate $\delta Q=\Gamma_M \left(\rho^H_{_M} - \rho_{_M}\right)\approx 0$ is very small. 
Dark energy is almost constant in time $\dot\rho_{_\Lambda}\approx 0$. It is the case for the matter-dominated episode in reheating, see Sec.~\ref{Msec}, and for stable massive pairs (cold dark matter) evolution, see Sec.~\ref{coldm}. 
\item In case (iii), $\delta Q\approx 0$ has two possibilities: (a) $\delta Q\gtrsim 0$ and $\rho^H_{_M}\gtrsim \rho_{_M}$, dark energy slowly converts to matter $\dot\rho_{_\Lambda}\lesssim 0$; (b) $\delta Q\lesssim 0$ and $\rho^H_{_M}\lesssim \rho_{_M}$, matter slowly converts to dark energy $\dot\rho_{_\Lambda}\gtrsim 0$. The (b) is the case for epochs after reheating. Dark energy converts to matter and reduces to its minimal value in reheating, and 
matter and radiation become dominant over (much larger than) dark energy. However, dark energy weakly couples to matter and radiation, i.e., $\delta Q\approx 0$ and $\dot\rho_{_\Lambda}\approx 0$. Its variation is much more slowly than matter/radiation decrease. Then it dominates over matter and radiation today. We present the preliminary discussions in Ref.~\cite{Xue2022}.
\end{enumerate} 
The inclusion of decay terms $\Gamma_M^{^{\rm de}}\rho_{_M}$ and transitions from one case to another are complex and need numerical studies.

\subsection{Unstable massive pair decay and reheating equation}\label{heat}

Coming from massive unstable pairs' decay, the radiation energy density $\rho_{_R}$ of relativistic particles $\bar\ell\ell$ (\ref{Mdecayr}) obeys the energy conservation law, see for example Ref.~\cite{Kolb1990},
\begin{eqnarray}
d(a^3\rho_{_R}) &=& -p_{_R} d(a^3) -d(a^3\rho_{_M})\nonumber\\ 
&=& -\frac{\rho_{_R}}{3} d(a^3) + (a^3\rho_{_M}) \Gamma^{^{\rm de}}_M dt,
\label{eqrho0}
\end{eqnarray}
where $d(a^3\rho_{_M})=-(a^3\rho_{_M}) \Gamma^{^{\rm de}}_M dt$ is the massive pair energy,  that converts to radiation energy. It leads to the reheating equation 
\begin{eqnarray}
\dot \rho_{_R} + 4 H \rho_{_R} = \Gamma^{^{\rm de}}_M\rho_{_M}.
\label{eqrho}
\end{eqnarray}
As a result, we have a close set of four ordinary differential
equations to uniquely determine 
the time evolution of the Hubble rate $H$, dark-energy density
$\rho_{_\Lambda}$, massive particles' energy density 
$\rho_{_M}$ and relativistic particles' energy density $\rho_{_R}$. They are generalised Friedman  equations (\ref{sfriedman0},\ref{sfriedman}) for $H$ and $\rho_{_\Lambda}$, the cosmic rate equation (\ref{rateeqd}) for $\rho_{_M}$, and 
the reheating equation (\ref{eqrho}) for $\rho_{_R}$. 
In addition, there are four algebraic relations: the massive pair plasma density $\rho^H_{_M}$ (\ref{apdenm}),
the pair-production rate 
$\Gamma_M$ (\ref{prate}), the Universe evolution $\epsilon$-rate (\ref{dde0}) 
and the pair-decay rate $\Gamma^{^{\rm de}}_M$ (\ref{Mdecayr}). We will numerically solve these equations, provided initial conditions are known. 

\begin{figure}[t]
\centering
\includegraphics[height=6.0cm,width=9.8cm]{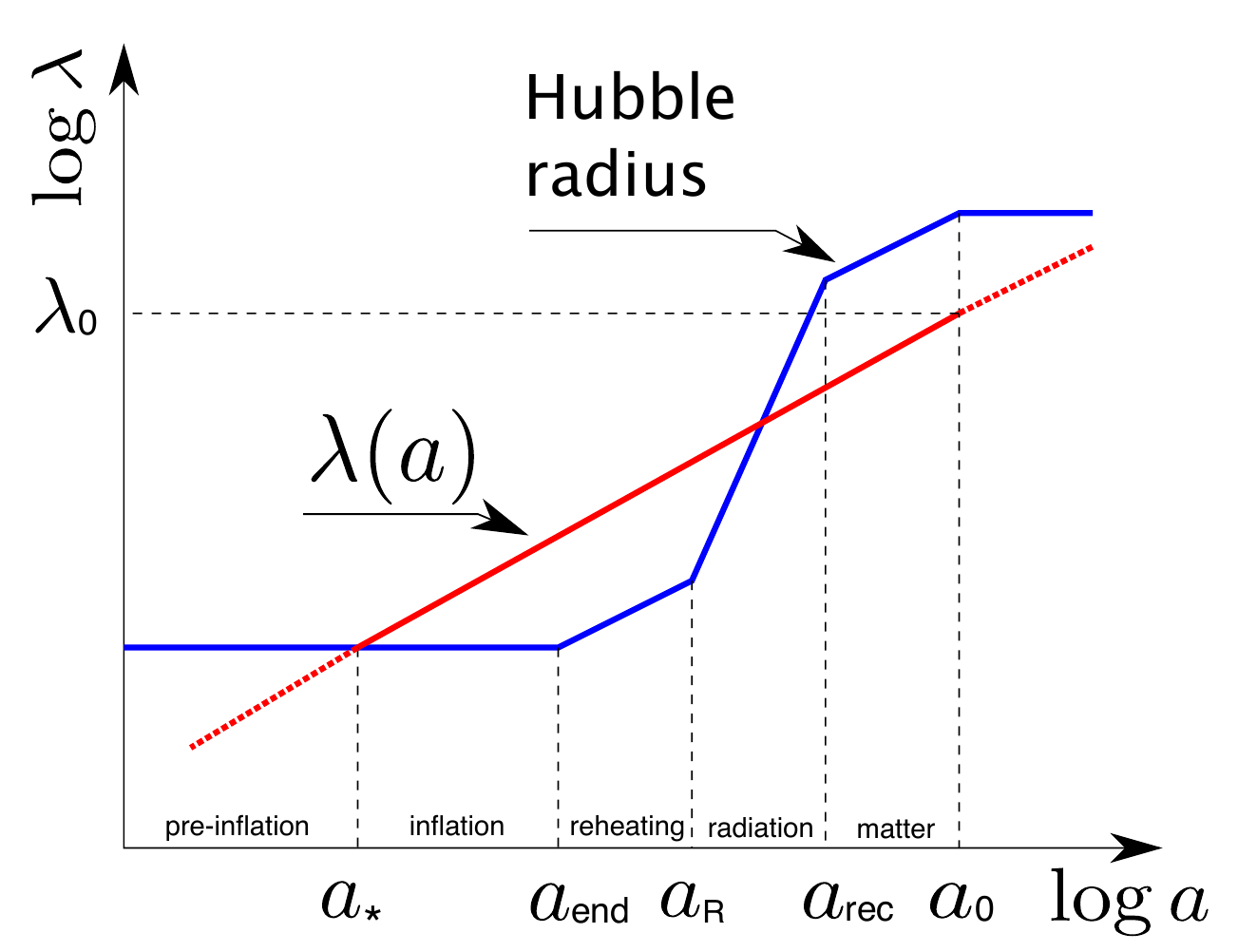}
\caption{We make this figure by modifying Fig.~1 in Ref.~\cite{Mielczarek2011}. 
Schematic evolution of the Hubble radius $H^{-1}$ and the physical length 
scale $\lambda(a)$, where physically interested scale 
$\lambda_0=\lambda(a_0)$ at the present time $a_0=1$ crossed the Hubble horizon $H_*$
at the early time $a_*$, fixed by the CMB pivot scale $\lambda_0=\lambda_*=k_*^{-1}$. The pre-inflation $a>a_*$, the inflation 
$a_*<a<a_{\rm end}$, 
the reheating $a_{\rm end}<a<a_{_R}$, 
and the recombination at $a_{\rm rec}$. 
}
\label{schematicf}
\end{figure} 

\section{Initial conditions and basic equations for reheating}\label{infend}

The inflation epoch ($H>\Gamma_{_M}$) ends, and the reheating epoch ($H<\Gamma_{_M}$) starts. The transitioning process must be very complex due to 
the back reactions of microscopic and macroscopic processes. We assume the transition to be instantaneous at the inflation end $a_{\rm end}$ and $H_{\rm end}$ when $H\lesssim \Gamma_{_M}$. For the inflation epoch from $a_*$ to $a_{\rm end}$, see Figure \ref{schematicf},
we obtain \cite{Xue2021}
\begin{eqnarray}
H_{\rm end}&=&H_*e^{-\epsilon_* N_{\rm end}}, \quad \epsilon_*=\chi (m_*/m_{\rm pl})^2,\quad 
\Delta_3\equiv 
(a_{\rm end}/a_*)=e^{N_{\rm end}}
\label{Hend}
\end{eqnarray}
where the inflation scale $H_*$ and $a_*$ correspond to the pivot scale $k_*=0.\,05\, ({\rm Mpc})^{-1}$ crossed the horizon $(
k_*=H_*a_*)$ for CMB observations \cite{Aghanim2020}. The observed spectral index $n_s$ and scalar amplitude $A_s$ determine $H_*=3.15\times 10^{-5}\,(r/0.1)^{1/2}m_{\rm pl}$ and the $\epsilon_*=(1-n_s)/2\approx 0.0175$ of the $\epsilon$-rate (\ref{dde0}) in inflation.
The $e$-folding number $N_{\rm end}$ and the 
tensor-to-scalar ratio 
$r$ are related by $H_{\rm end}\lesssim \Gamma_M$ 
\begin{eqnarray}
r\lesssim  
7.97\times 10^4 \chi  (1-n_s)^3e^{(1-n_s) N_{\rm end}},
\label{endr}
\end{eqnarray}
and $\chi> 0$ implies $r>0$. The observational constraint on the tensor-to-scalar ratio and spectra index $(r,n_s)$, see Fig.~2 of Ref.~\cite{Xue2021}, gives $\chi\lesssim {\mathcal O}(10^{-3})$. 
The small $H$ variation implies 
at the inflation end 
\begin{eqnarray}
H^2_{\rm end} = \frac{\rho^{\rm end}_{_\Lambda}  + \rho^{\rm end}_{_M}}{3m^2_{\rm pl}}\approx\frac{\rho^{\rm end}_{_\Lambda}}{3m^2_{\rm pl}}; \quad 
\rho^{\rm end}_{_\Lambda}\gg \rho^{\rm end}_{_M},
\label{ratioend}
\end{eqnarray}
and $\rho^{\rm end}_{_\Lambda}\approx \rho^{\rm end}_c\equiv 3
m^2_{\rm pl}H^2_{\rm end}$. 
We approximately adopt the value 
\begin{eqnarray}
\Omega^{\rm end}_{_M} = \rho^{\rm end}_{_M}/ \rho^{\rm end}_c \approx 4.7\times 10^{-3},
\label{ipe0}
\end{eqnarray}
for which $(\Gamma_M/H)_{\rm end}\approx 1$. These are the reheating epoch initial conditions. 

Using the characteristic scale $H_{\rm end}$ and
density $\rho_c^{\rm end}$, we normalize 
\comment{
In the inflation epoch $a_*<a<a_{\rm end}$, the scale factor increases from the beginning $a_*$ to the end $a_{\rm end}$ 
\begin{eqnarray}
\Delta_3\equiv \frac{a_3}{a_4}=\frac{a_{\rm end}}{a_*}=e^{N_{\rm end}}, 
\label{d3}
\end{eqnarray}
where $N_{\rm end}$ is the $e$-folding number, 
$a_*$ and $H_*$ correspond to the interesting mode of the pivot length scale $k_*=a_*H_*$ crossed the horizon 
for CMB power spectra observations. 
}
$h\equiv H/H_{\rm end}$, 
\begin{eqnarray}
\Omega_{_{\Lambda,M,R}}\equiv\frac{ \rho_{_{\Lambda,M,R}}}{\rho_c^{\rm end}}, \quad \Omega^H_{_M}\equiv \frac{\rho^H_{_M}}{\rho_c^{\rm end}}=\frac{2}{3}\chi (\hat m/m_{\rm pl})^2h^2.
\label{ipe}
\end{eqnarray}
Here we introduce the mass parameter $\hat m$ or $\chi (\hat m/m_{\rm pl})^2$ as a typical scale parameter for reheating and will fix its value by observations.
Thus, we recast the Friedman equations  (\ref{sfriedman0}) and (\ref{sfriedman}), the cosmic rate equation (\ref{rateeqd}) and reheating equation (\ref{eqrho}) as,
\begin{eqnarray}
h^2 &=& \Omega_{_\Lambda}  + \Omega_{_M} + \Omega_{_R},
\label{eqhp}\\
\frac{dh^2}{dx} &\approx & -3\Omega_{_M}-4\Omega_{_R},
\label{rehp}\\
\frac{d\Omega_{_M}}{dx}+ 3\Omega_{_M} & = & \frac{\Gamma_M}{H} \left(\Omega_{_M}^H - 
\Omega_{_M}\right) -\frac{\Gamma^{^{\rm de}}_M}{H}
\Omega_{_M}, \label{rateeqp}\\
\frac{d\Omega_{_R}}{dx} + 4 \Omega_{_R} &=& \frac{\Gamma^{^{\rm de}}_M}{H}\Omega_{_M}.
\label{reheateq}
\end{eqnarray}
The ratios are
\begin{eqnarray}
\frac{\Gamma_M}{H} =\left(\frac{\chi}{4\pi}\right)\left(\frac{\hat m}{ H_{\rm end}}\right)
\frac{\epsilon}{h};\quad
\frac{\Gamma^{^{\rm de}}_M}{H} =g^2_{_Y}\left(\frac{ \hat m}{H_{\rm end}}\right)\frac{1}{h},
\label{rohm}
\end{eqnarray}
and the $\epsilon$-rate (\ref{dde0}) becomes 
\begin{eqnarray}
\epsilon &\equiv & -\frac{1}{H}\frac{d H}{dx}
= \frac{3}{2}\frac{\Omega_{_M}+(4/3)\Omega_{_R}}{\Omega_{_\Lambda}+\Omega_{_M}
+\Omega_{_R}}.
\label{dde}
\end{eqnarray}
Instead of the cosmic time $t$, here we adopt the cosmic $e$-folding variable $x=\ln (a/a_{\rm end})$ 
and $d(\cdot\cdot\cdot)/dx=d(\cdot\cdot\cdot)/(Hdt)$
for the sake of simplicity and significance in physics.
In the next sections, we will numerically integrate these basic equations (\ref{eqhp}-\ref{dde}) for the reheating epoch by using the inflation end (\ref{ipe0}) as the initial condition.

We have to emphasize that the differential equations (\ref{eqhp}-\ref{reheateq}) represent a macroscopic back-and-forth reaction system characterized by the scales $\tau_{_H}$, $\tau_{_M}$ and $\tau_{_R}$. It describes the processes: inflation, reheating and standard cosmology. 
It differs from the differential equations (\ref{ffriedman0}-\ref{ffriedman}) and those in Sec.~\ref{qppo} for a microscopic back-and-forth reaction system of fast-oscillating components characterized by the scale ${\mathcal O}(1/M)$. The fast components' contributions are effectively represented by the massive pair plasma state $\rho^H_{_M}$ (\ref{apdenm}) that enters the cosmic rate equation (\ref{rateeqp}).  

\section{\bf Different episodes in reheating epoch}
\label{reheatingdetails}
  
In the reheating epoch, generally speaking, the horizon $h$ and 
the dark energy $\Omega_{_\Lambda}$ 
decreases, as the matter content $\Omega_{_M}$ or $\Omega_{_R}$ increases, meanwhile
the ratio $\Gamma_M/H$ (\ref{rohm}) and the $\epsilon$-rate (\ref{dde}) increase.
To gain insight into the physics first, we use the $\epsilon$-rate 
values (\ref{dde}) to characterize the different episodes 
in the reheating epoch. In each episode, the $\epsilon$ rate slowly varies in time, 
we approximately have the time scale of the spacetime expansion $H^{-1}\approx  \epsilon t$.
In the transition from one episode to another, the $\epsilon$-rate 
significantly changes its value. Using the characteristic $\epsilon$ values
$\epsilon\ll 1, \epsilon\approx 3/2,\epsilon \approx 2$, 
we identify the following three different episodes ${\mathcal P}$-{\it episode},  
${\mathcal M}$-{\it episode} and  ${\mathcal R}$-{\it episode} 
in the reheating epoch. 
 
\comment{where the Hubble radius $H^{-1}$ represents the horizon for causal microphysics on distances less than 
the Hubble radius, as the Hubble radius is the distance 
a light signal can travel in an expansion time $t$.}

\subsection{Preheating ${\mathcal P}$-{\it episode}: dark energy $\rho_{_\Lambda}$ converting into matter $\rho_{_M}$}\label{pre}

The short preheating ${\mathcal P}$-{\it episode} is the transition from the inflation end to the reheating start. 
In this episode, the pair production rate $\Gamma_M$ (\ref{prate}) 
is larger than the Hubble rate $H$, that is still much larger than 
the pair decay rate $\Gamma^{^{\rm de}}_M$ (\ref{Mdecayr}),
\begin{eqnarray}
\Gamma_M >H \gg \Gamma^{^{\rm de}}_M, \quad \rho_{_\Lambda}>\rho^H_{_M}>\rho_{_M}\gg \rho_{_R}.
\label{Mdecayp} 
\end{eqnarray}
The radiation energy density is completely 
negligible $\rho_{_R}\approx 0$. 
We neglect massive pairs decay to light particles $\Gamma^{^{\rm de}}_M\approx 0$ (\ref{Mdecayr}).
The reheating equation (\ref{reheateq}) is then not relevant, and the 
basic equations (\ref{eqhp}), (\ref{rehp}) and (\ref{rateeqp}) reduce to
\begin{eqnarray}
h^2 &=& \Omega_{_\Lambda}  + \Omega_{_M},
\label{eqhpm}\\
dh^2/dx &=& -3\Omega_{_M},
\label{rehpm}\\
d\Omega_{_M}/dx+ 3\Omega_{_M} & = & (\Gamma_M/H) \left(\Omega_{_M}^H - 
\Omega_{_M}\right), \label{rateeqpr} 
\end{eqnarray}
where the ratio $\Gamma_M/H>1$ (\ref{rohm}) increases as the $\epsilon$-rate (\ref{dde})
\begin{eqnarray}
\epsilon &\approx &\frac{3}{2}\frac{\rho_{_M}}{\rho_{_\Lambda}+\rho_{_M}}=\frac{3}{2}\frac{\Omega_{_M}}{\Omega_{_\Lambda}+\Omega_{_M}}.
\label{ddep}
\end{eqnarray}
In the ${\mathcal P}$-{\it episode}, these equations uniquely determine the evolution of the Hubble rate $H$, 
pairs' energy densities $\rho_{_M}$ and dark-energy density $\rho_{_\Lambda}$.

\begin{figure*}[t]
\includegraphics[height=5.5cm,width=7.8cm]{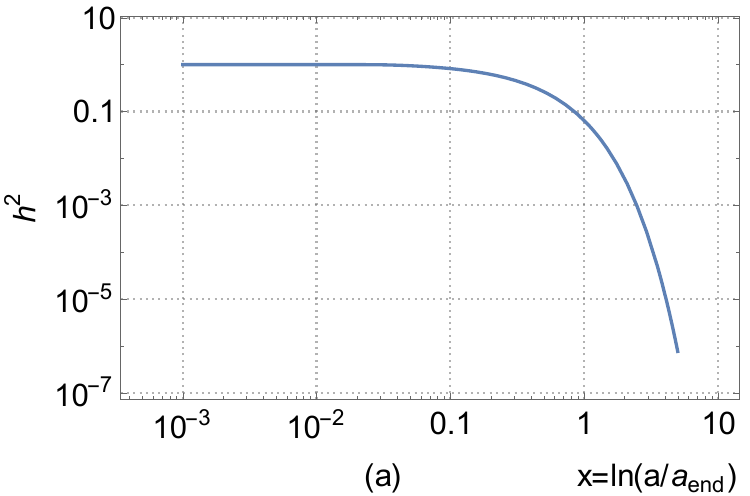}
\hspace{0.333cm}
\includegraphics[height=5.5cm,width=7.8cm]{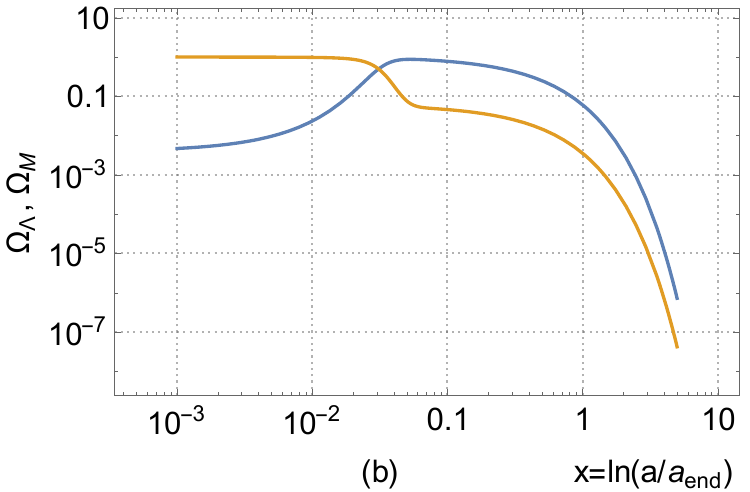}
\includegraphics[height=5.5cm,width=7.8cm]{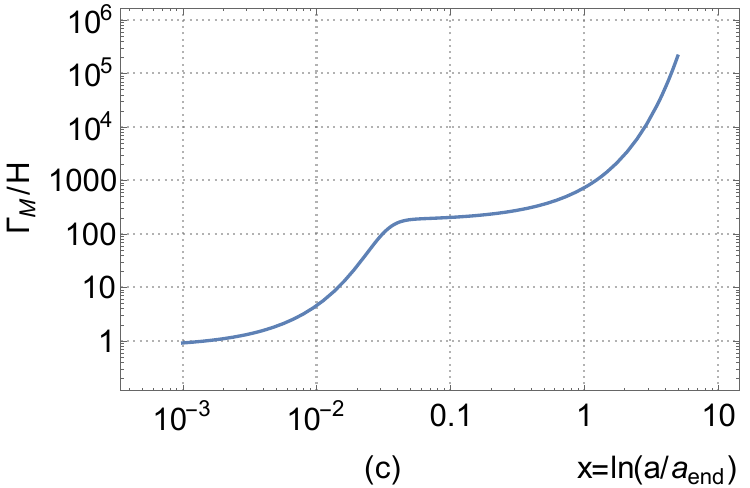}\hspace{0.333cm}
\includegraphics[height=5.5cm,width=7.8cm]{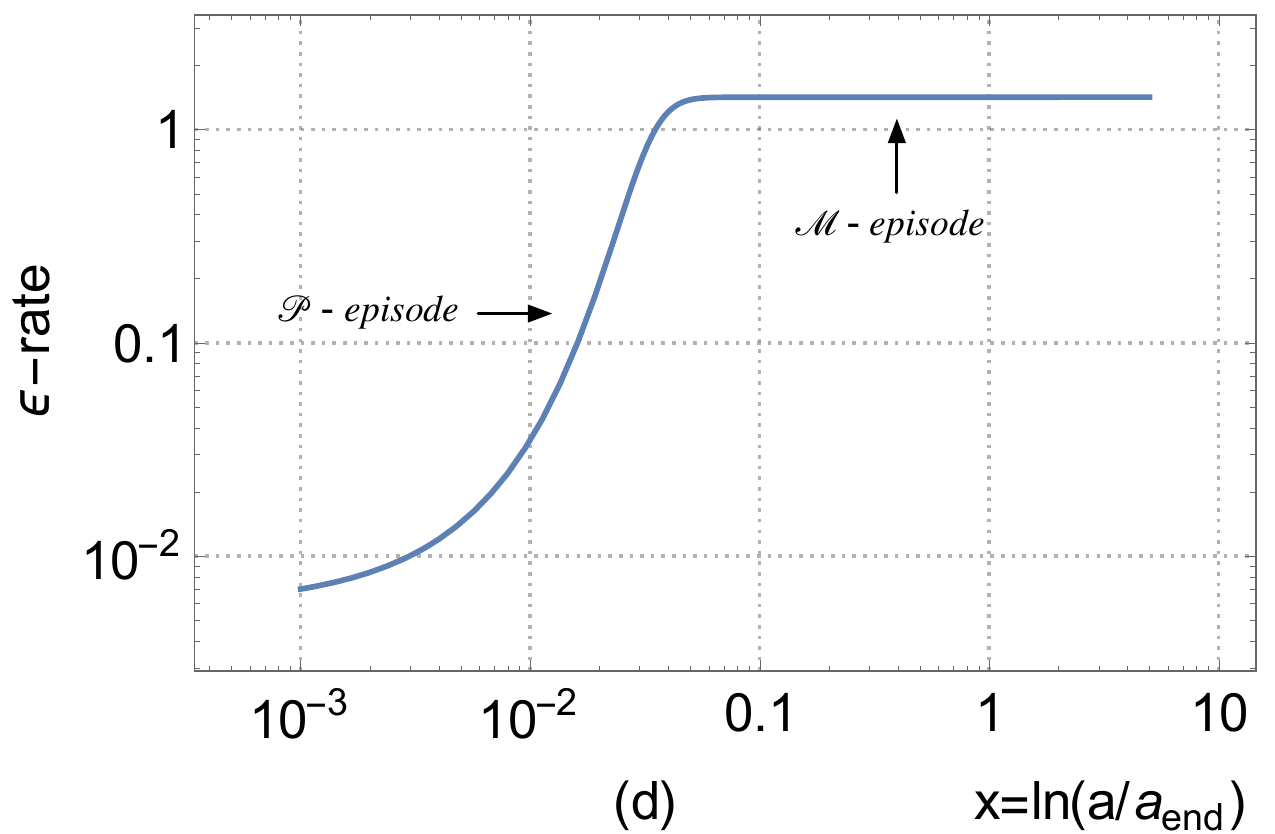}\hspace{0.333cm}
\caption{ (Color Online). In a few $e$-folding number $x=\ln(a/a_{\rm end})$, 
(a) the Hubble rate drops rapidly; (b) the massive pair energy density exceeds the dark-energy density; (c) the ratio $\Gamma_M/H>1$ increases rapidly; 
(d) the $\epsilon$-rate of $H$ variation increases in the transition 
from $\epsilon \ll 1$ (inflation epoch) to the asymptotic value
$\epsilon \sim {\mathcal O}(1)$ (${\mathcal M}$-{\it episode}, see Sec.~\ref{Msec}). 
We plot these solutions with the initial condition (\ref{ipe0})
and parameter $(\hat m/m_{\rm pl})=27.7$. 
}\label{AllBr}
\end{figure*}

\subsubsection{High efficiency of dark energy converting into matter}\label{prec}
   
\begin{figure*}[t]
\includegraphics[height=5.5cm,width=7.8cm]{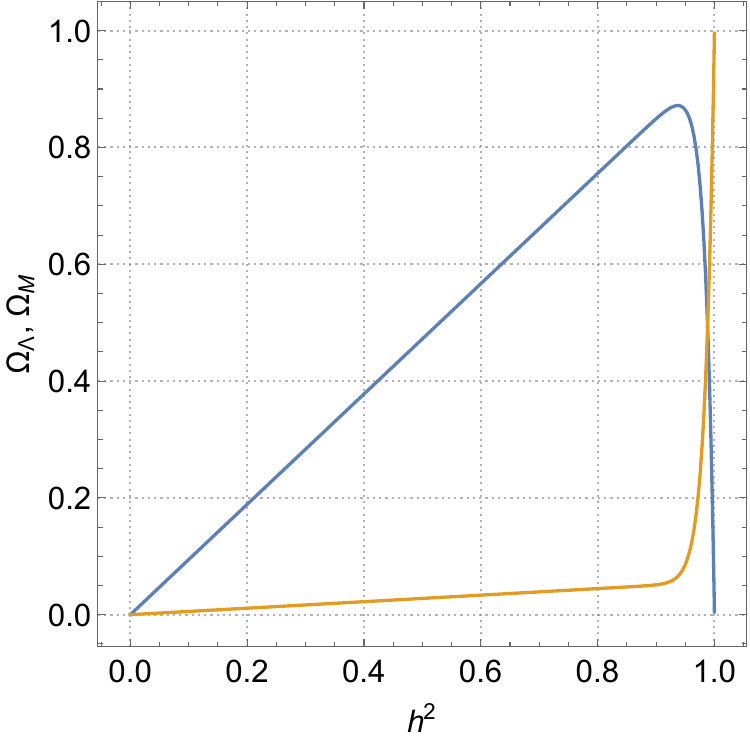} 
\includegraphics[height=5.5cm,width=7.8cm]{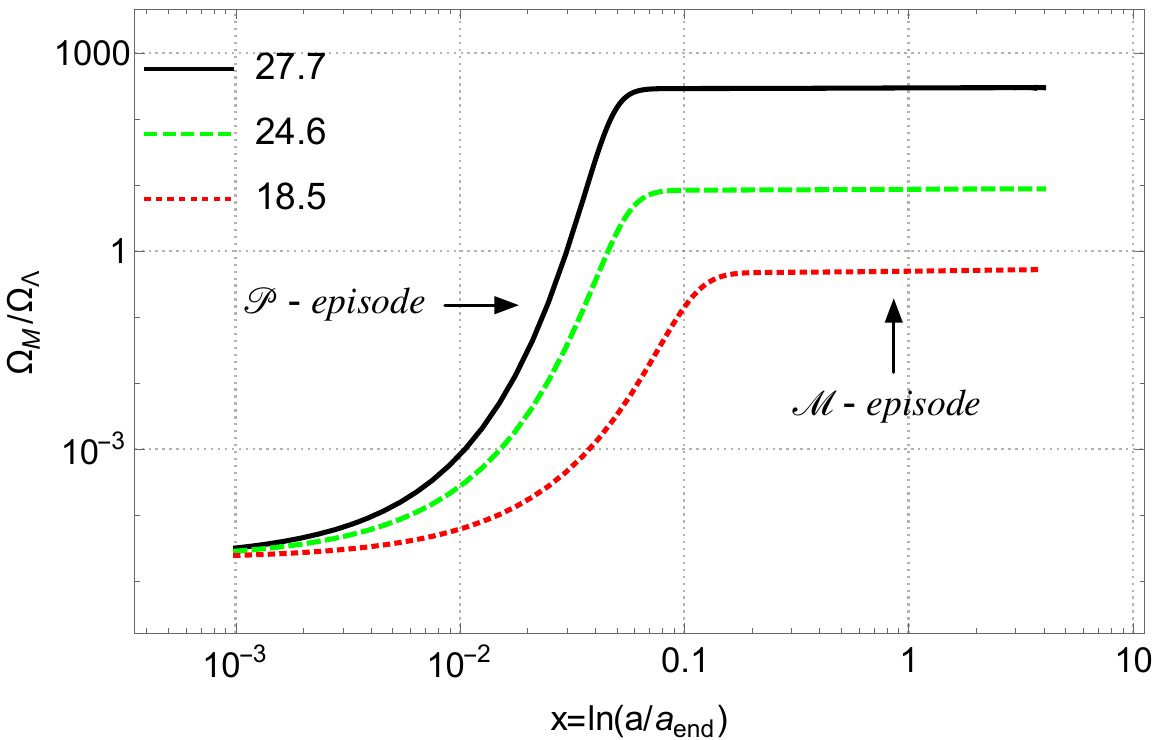} 
\caption{ (Color Online). Left: The energy densities 
$\Omega_{_\Lambda}$ (orange) and $\Omega_{_M}$ (blue) are plotted as 
functions of the horizon $h^2$, corresponding to Figs.~\ref{AllBr} (a) and (b). Right: The ratio $\Omega_{_M}/\Omega_{_\Lambda}=\rho_{_M}/\rho_{_\Lambda}$ varies in the transition from 
$\rho_{_M}/\rho_{_\Lambda}\ll 1$ (inflation epoch) 
to $\rho_{_M}/\rho_{_\Lambda}\gg 1$, approaching to a constant
(${\mathcal M}$-{\it episode}). 
We plot the ratio $\rho_{_M}/\rho_{_\Lambda}$ for selected values $\hat m/m_{\rm pl} =27.7,24.6,18.5$, corresponding 
to the solid black line, 
green dashed line, 
red dotted line. 
The initial conditions are $H_{\rm end}$ (\ref{Hend}) and $\Omega_M^{^{\rm end}}$ (\ref{ipe0}).
}\label{mscale}
\end{figure*}

Using the values $H_{\rm end}$ (\ref{Hend}) and $\Omega_M^{^{\rm end}}$ (\ref{ipe0}) at the inflation end as the initial conditions for the ${\mathcal P}$-episode, 
we numerically integrate 
Eqs.~(\ref{eqhp}), (\ref{rehp}) and (\ref{rateeqp}), by selecting values of the mass 
parameter $\hat m/m_{\rm pl}$. 
In Figs.~\ref{AllBr} and \ref{mscale}, the numerical solutions are plotted 
in terms of the $e$-folding variable 
$x=\ln(a/a_{\rm end})$. These solutions show 
an important result that the 
dark-energy density $\rho_{_\Lambda}$ 
is significantly converted to the matter-energy density $\rho_{_M}$, 
as the pair-production rate $\Gamma_M$ increases and becomes much larger than 
the Hubble rate $H$. In more detail,
we list that in the ${\mathcal P}$-{\it episode} the physical quantities 
vary in time as follows, 
\begin{enumerate}[(i)]
\item  the Hubble rate $h$ decreases rapidly in a few $e$-folding number, as $\rho_{_M}$ becomes dominate over $\rho_{_\Lambda}$, see 
Fig.~\ref{AllBr} (a);
\item the $\rho_{_M}$ increases at the expense of the $\rho_{_\Lambda}$, eventually $\rho_{_M}$ exceeds and dominates $\rho_{_\Lambda}$, see Fig.~\ref{AllBr} (b);
\item the ratio $\Gamma_M/H$ (\ref{Mdecayp}) increases and becomes much larger than unity ($\Gamma_M/H\gg 1$), see Fig.~\ref{AllBr} (c);
\item the $H$ varying rate $\epsilon$
(\ref{ddep}) increases from $\epsilon \ll 1$ to $\epsilon\sim {\mathcal O}(1)$, indicating the transition from the inflation end 
to the preheating ${\mathcal P}$-episode, and it then 
approaches to an asymptotic value, see 
Fig.~\ref{AllBr} (d).
\end{enumerate}

In Figure \ref{mscale} (left), we plot the energy densities 
$\Omega_{_\Lambda}$ and $\Omega_{_M}$ as 
functions of the horizon $h^2$, corresponding to Figures (a)
and (b) in Fig.~\ref{AllBr}.  It shows two branches of asymptotic solutions
$\Omega_{_\Lambda}$ (orange) respectively, 
\begin{eqnarray}
\Omega_{_\Lambda} \simeq \alpha^1_{_\Lambda} h^2 ~~(h^2< 0.95);\quad
\Omega_{_\Lambda} \simeq   \alpha^2_{_\Lambda} h^2 ~~ (h^2> 0.95);
\quad \alpha^2_{_\Lambda} \gg\alpha^1_{_\Lambda},
\label{largeh}
\end{eqnarray}
$\Omega_{_M}+\Omega_{_\Lambda}=h^2$ and the turning point is 
about $h^2\approx 0.98$, at which $\Omega_{_M}$ exceeds 
$\Omega_{_\Lambda}$ and the rapid $\rho_{_\Lambda}\Rightarrow \rho_{_M}$ 
converting process takes place. 
The characteristic  behaviour $\Omega_{_\Lambda} \propto h^2$ (\ref{largeh}) is the same as that
in the pre-inflation and inflation epochs.
Correspondingly, two branches of asymptotic 
solutions for the matter $\Omega_{_M}$ (blue) 
are 
\begin{eqnarray}
\Omega_{_M} \simeq \alpha^1_{_M} h^2 ~(h^2< 0.95);\quad
\Omega_{_M} \simeq \Omega^{\rm max}_{_M} - \alpha^2_{_M} (h^2-0.95) ~ (h^2> 0.95),
\label{largehm}
\end{eqnarray}
and $\alpha^2_{_M} \gg\alpha^1_{_M}$, where $\Omega^{\rm max}_{_M}\approx 0.85$ when $h^2\approx 0.95$. The 
coefficients $\alpha^{1,2}_{_\Lambda}$ and $\alpha^{1,2}_{_M}$ 
in Eqs.~(\ref{largeh}) and (\ref{largehm}) can be numerically obtained. As $h^2\rightarrow 0$ and $(a/a_{\rm end})$ increases, 
$\Omega_{_M}\rightarrow 0$, $\Omega_{_\Lambda}\rightarrow 0$ 
and $\Omega_{_M}\gg \Omega_{_\Lambda} (\alpha^1_{_M}\gg \alpha^1_{_\Lambda})$. 

Figure \ref{mscale} (right) shows that in the preheating ${\mathcal P}$-episode, the dark energy density $\rho_{_\Lambda}$ converts to the matter-energy $\rho_{_M}$.
The ratio $\rho_{_M}/\rho_{_\Lambda}$ rapidly 
increases in a few $e$-folding numbers from $\rho_{_M}/\rho_{_\Lambda}\ll 1$ 
at the inflation end to a value 
$\rho_{_M}/\rho_{_\Lambda}\gtrsim {\mathcal O}(1)$. Then $\rho_{_M}$ becomes dominant. 
The Hubble rate $H$ rapidly decreases and becomes much smaller than the pair-production rate $\Gamma_M$.  
We define the ${\mathcal P}$-episode end at $1.11 a_{\rm end}$ by $\rho_{_\Lambda}\approx \rho_{_M}$
from Figs.~\ref{AllBr} and \ref{mscale} (right). 
It shows
that the preheating ${\mathcal P}$-episode is a very brief transition episode.

We ought to discuss the condition for high efficiency of dark energy converting into massive pairs' energy in this preheating ${\mathcal P}$-{\it episode}. 
Figure \ref{AllBr} (b) shows $\rho_{_\Lambda}$ decreases and $\rho_{_M}$ increases rapidly,  the efficiency of dark energy converting matter-energy is large. The reasons are the following. (i) The ratio $\Gamma_M/H\gg 1$ increases rapidly as $H$ decreases rapidly, see Fig.~\ref{AllBr} (c) and (a), respectively. 
(ii) The $\epsilon$ rate (\ref{dde0}) rapidly increases to the order of unity, see Fig.~\ref{AllBr} (d). (iii) The mass parameter $\hat m$ (\ref{prate}) is large, see Fig.~\ref{mscale} (right), 
which implies heavy mass $M$ and large number ${\mathcal N}_{\rm pair}$ of pairs produced in the massive pair plasma state (\ref{apdenm}). 
If the mass parameter $\hat m$ is small, the dark energy and matter conversion efficiency is small, 
see the ratio $\Omega_M/\Omega_\Lambda$ in Fig.~\ref{mscale} (right).  
{\it A priori}, we do not have theoretical arguments about how much $\hat m$ value is. Instead, we select $\hat m$ values {\it a posteriori} in comparison with observations, and the Universe does not stay cold state of $\rho_{_\Lambda}>\rho_{_M}$. The necessary condition is the existence of a threshold $\hat m_{\rm thresh}$ for which  $\rho_{_M}>\rho_{_\Lambda}$ of efficient conversion in reheating.

\subsubsection{Threshold of massive pair mass and number for $\rho_{_M}>\rho_{_\Lambda}$ }\label{threshold}

Figure \ref{mscale} (right) shows
an important result. These solutions depend on the 
pair mass parameter $\hat m$ (\ref{apdenm}) 
introduced for the reheating epoch. There exists a theoretical threshold on the mass parameter $\hat m_{\rm thresh}\approx 20 m_{\rm pl}$. 
\begin{enumerate}[(a)]
\item For large mass parameters 
$\hat m/m_{\rm pl} >20$, 
the pair energy density $\rho_{_M}$ exceeds the dark-energy density 
$\rho_{_\Lambda}$ and the asymptotic value $\rho_{_M}/\rho_{_\Lambda}> 1$,
\begin{eqnarray}
\hat m> \hat m_{\rm thresh} \approx 20 m_{\rm pl},\quad \rho_{_M}/\rho_{_\Lambda}> 1.
\label{mthre}
\end{eqnarray}
The reason is that the number ${\mathcal N}_{\rm pair}$ (or effective degeneracy $g_d$) of 
massive pairs produced in the reheating epoch has to be large enough so that
$\Gamma_M\gg H$ and the conversion from $\rho_{_\Lambda}$ to $\rho_{_M}$ is 
efficient. It corresponds to the physical situation that the most radiation and matter of the Universe is generated in
the reheating epoch. 
\item For small mass parameters $\hat m/m_{\rm pl} <  20$, 
the massive pairs' energy density  
$\rho_{_M}$ never exceeds the dark-energy density $\rho_{_\Lambda}$, namely $\rho_{_M}/\rho_{_\Lambda} < 1$. The conversion from $\rho_{_\Lambda}$ to $\rho_{_M}$ is 
inefficient.
This case corresponds to the unrealistic situation that the Universe 
inflation would never have completely ended, i.e., 
the cosmological term $\rho_{_\Lambda}$ always dominates $H^2$.
\end{enumerate}

Observe that the mass parameter $\hat m$ of the reheating epoch 
is larger than the mass parameter $m_*$ of the inflation epoch. From the viewpoint of pair production, 
the pair mass scale in reheating should be smaller than that in inflation since the reheating horizon $H$ 
is smaller than the inflation one. 
Therefore, it implies that the effective 
numbers ${\mathcal N}_{\rm pair}$
of massive pairs produced in reheating, $\Gamma_M/H > 1$ is much larger than  
that of massive pairs produced in inflation $\Gamma_M/H < 1$. These massive pairs contain both stable and unstable pairs.

Equation (\ref{ddep}) shows that the asymptotic value of the Horizon variation
$\epsilon$-rate (\ref{dde0}) relates to the ratio $\rho_{_M}/\rho_{_\Lambda}$ 
asymptotic value, see Figs.~\ref{AllBr} (d) and \ref{mscale} (right).   
For large mass parameter $\hat m/m_{\rm pl} \gtrsim 27.7$, 
the ratio $\rho_{_M}/\rho_{_\Lambda} \gg 1$ \footnote{This condition also admits the possibility of a small negative dark energy density $\rho_{_\Lambda}<0$ and $|\rho_{_\Lambda}|\ll \rho_{_{M}}$. Namely, Figure \ref{AllBr} (b) admits solution $\rho_{_{M}}\gg |\rho_{_\Lambda}|$ and $\rho_{_\Lambda}$ drops slightly below zero.}, 
the $\epsilon$-rate (\ref{ddep}) approaches to the asymptotic value 
$\epsilon\approx \epsilon_{_M}=3/2$. It shows the episode of 
massive pairs domination: ${\mathcal M}$-{\it episode}. 

\subsubsection{Minimal comoving radius $(Ha)^{-1}$ location} 

Before discussing the ${\mathcal M}$-{\it episode}, we would like to mention the turning point at which the Universe acceleration vanishes $\ddot a=0$,
\begin{eqnarray}
2\rho_{_\Lambda}
=(1+3\, \omega_{_M})\,\rho_{_M}-\!(1\!+\!3\omega_{_R})\rho_{_R},
\label{ustop}
\end{eqnarray}
which is obtained from the $1\!-\!1$ component of the Einstein equation 
\begin{eqnarray}
2\frac{dH}{dt}\! +\! 2H^2 \!=\frac{2\ddot a}{ a}=\!
\Big[2\rho_{_\Lambda}\!-\!(1\!+\!3\omega_{_M})\rho_{_M}-\!(1\!+\!3\omega_{_R})\rho_{_R}\Big].
\label{acc0}
\end{eqnarray} 
At this turning point, the Universe stops acceleration $\ddot a >0$ 
and starts deceleration $\ddot a <0$. The turning point occurs at 
$\rho_{_\Lambda}=\rho_{_M}/2$ for $\omega_{_M}\approx 0$ and $\rho_{_R}\approx 0$.
It tells us the balance point
of the competition between $\rho_{_\Lambda}$ 
and $\rho_{_M}$ in the ${\mathcal P}$-{\it episode}. 

On the other hand, the minimal value of the comoving radius $(aH)^{-1}$ locates at
\begin{eqnarray}
d(aH)^{-1}/dt =0\quad \Rightarrow \quad \dot H+H^2=0.
\label{minc0}
\end{eqnarray} 
From Friedman equations (\ref{friedman}), 
we obtain 
\begin{eqnarray}
\rho_{_\Lambda}=\rho_{_M}/2 +\rho_{_R}\approx \rho_{_M}/2,\quad \epsilon=\epsilon^{\rm min}_{_\Lambda} =1,
\label{minc}
\end{eqnarray}
coinciding with the turning point (\ref{ustop}). 
Namely at the minimal comoving radius $(aH)^{-1}$,
the Universe stops acceleration $\ddot a >0$ and begins deceleration 
$\ddot a <0$, starting the reheating epoch and standard cosmology.  
This is indeed the case for large mass parameter $(\hat m/m_{\rm pl}) >  20$ and 
the ratio $\rho_{_M}/\rho_{_\Lambda}$ becomes larger than 2. 
The numerical results (Fig.~\ref{AllBr}) show that this turning/minimal point 
$\epsilon_{\rm min}=1$ locates at 
$x_{\rm min}\approx 1.7\times 10^{-2}$ and $a_{\rm min}\approx a_{\rm end}\times \exp ~(1.7\times 10^{-2})=1.02~ a_{\rm end}$. 

While the turning/minimal point $\epsilon_{\rm min}=1$ is never reached, for the cases of the small mass parameter $(\hat m/m_{\rm pl}) <  20$ and the ratio $\rho_{_M}/\rho_{_\Lambda}$ is always smaller than 2, see Fig.~\ref{mscale} (right). The reason is that dark energy converting to matter is inefficient, the massive pairs energy is not large enough to balance the dark energy and slow down the Universe's acceleration.
The Universe keeps acceleration $\ddot a >0$ and does not run into the reheating epoch.
Therefore, the mass parameter range below the threshold $\hat m_{\rm thresh}$ (\ref{mthre}) $(\hat m/m_{\rm pl}) <  20$ should be excluded. 

\subsection{Massive pairs domination: ${\mathcal M}$-{\it episode}}\label{Msec}

After the ${\mathcal P}$-{\it episode} transition,
it is the $\mathcal M$-{\it episode} of massive pair domination characterised by 
\begin{eqnarray}
\rho_{_M}&\gg& \rho_{_\Lambda}\gg \rho_{_R},\quad
\Gamma_M > H > \Gamma^{^{\rm de}}_M.
\label{epm0}
\end{eqnarray}
The radiation energy density $\rho_{_R}$ is negligible in the basic 
equations (\ref{eqhp}-\ref{dde}). The $H$ variation $\epsilon$-rate $\epsilon_{_M} \approx  3/2$ in Fig.~\ref{AllBr} (d) for 
$\rho_{_M}/\rho_{_\Lambda}\gg 1$ in 
Fig.~\ref{mscale} (right). In this episode, 
the Hubble rate $H$ and scale factor $a(t)$ vary as
\begin{eqnarray}
H^{-1}\approx \epsilon_{_M} t,
\quad 
a(t)\sim t^{1/\epsilon_{_M}},
\label{epm}
\end{eqnarray}
$h^2\approx \Omega_M$ and the pair energy density 
$\Omega_M\propto (a/a_{\rm end})^{-2\epsilon_{_M}}$.


\begin{figure*}[t]
\includegraphics[height=5.5cm,width=7.8cm]{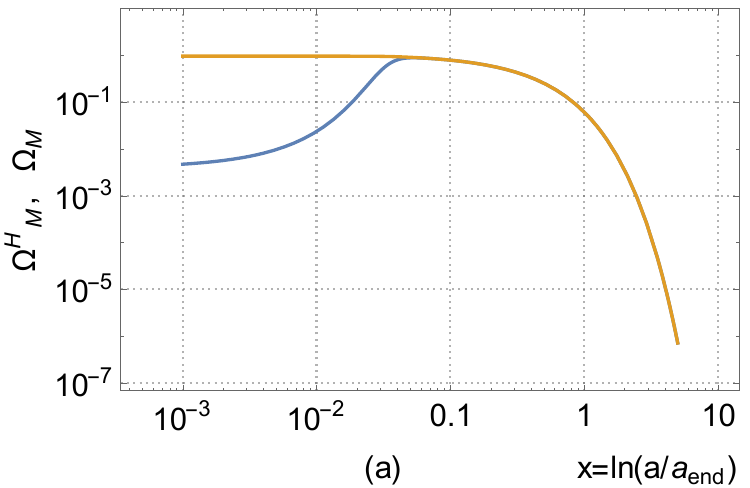}
\hspace{0.333cm}
\includegraphics[height=5.5cm,width=7.8cm]{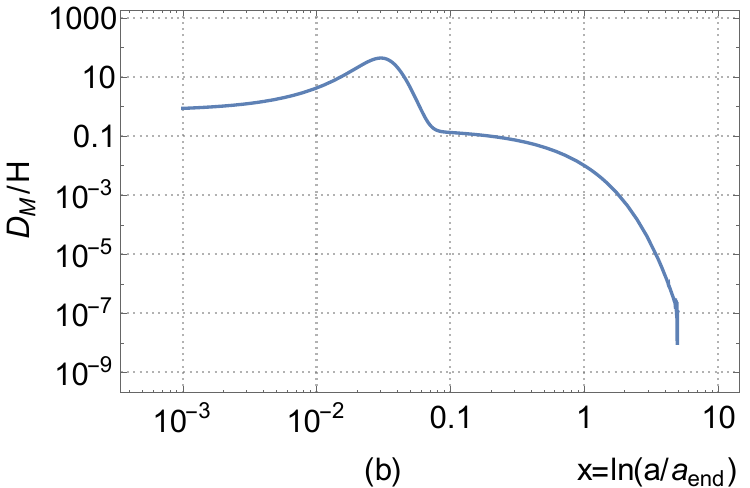}
\caption{(Color Online). In a few $e$-folding number $x=\ln(a/a_{\rm end})$, 
(a) the energy density $\rho^H_{_M}$ (\ref{ipe}) (orange) 
and the solution $\rho_{_M}$ (blue) to 
the cosmic rate equation (\ref{rateeqpr}), showing $\rho_{_M}$ approaches $\rho^H_{_M}$; 
(b) the detailed balance term $D_M$ (\ref{dbt}) vanishes.
We plot these solutions with the initial condition (\ref{ipe0})
and parameter $(\hat m/m_{\rm pl})=27.7$. 
}\label{balance}
\end{figure*} 

Moreover, the back-and-forth processes (\ref{bfrho}) are important, as described by   
the cosmic rate equation (\ref{rateeqpr}) with the detailed balance term $D_M$,
\begin{eqnarray}
D_M\equiv \Gamma_M \left(\Omega_{_M}^H - \Omega_{_M}\right),
\label{dbt}
\end{eqnarray}
and we define its characteristic time scale $\tau_{_D}$ 
\begin{eqnarray}
\tau^{-1}_{_D}\equiv \frac{D_M}{\Omega_{_M}}=\frac{\Gamma_M}{\Omega_{_M}} \left(\Omega_{_M}^H - \Omega_{_M}\right).
\label{dbtc}
\end{eqnarray} 
Note that $\tau_{_D}$ differs from $\tau_{_M}=\Gamma_M^{-1}$ (\ref{prate}).
The microscopic time scale $\tau_{_M}$ is much smaller 
than the macroscopic expansion time scale $\tau_{_H}=H^{-1}$, 
$\tau_{_M}\ll \tau_{_H}$.
Therefore, the 
back-and-forth (\ref{bfrho}) can build 
an energy equipartition $\rho_{_M}\approx \rho^H_{_M}$, and 
the detailed balance term (\ref{dbt}) vanish in its time-averaged 
\begin{eqnarray}
\langle \rho_{_M}- \rho^H_{_M}\rangle =0,
\label{avede}
\end{eqnarray}
over the macroscopic time $\tau_{_H} \gg \tau_{_M}$. 
The cosmic rate equation becomes approximately  
\begin{eqnarray}
\frac{d\rho_{_M}}{dt}+ 3 H\rho_{_M} &\approx & 0,
\label{rateeq0}
\end{eqnarray}
whose solution is $\rho_{_M}\propto a^{-3}$. It is consistent 
with the matter-dominated solution to Eq.~(\ref{sfriedman}), 
yielding $H^2\sim \rho_{_M}\propto a^{-3}$. It is also self-consistent with
the pair plasma density (\ref{apdenm}) 
$\rho^H_{_M}=2\chi \hat m^2 H^2 \propto  a^{-3}$. 

In order to verify these discussions and $\rho_{_M}\approx \rho^H_{_M}$, we check the solution 
(\ref{avede}) or (\ref{rateeq0}) 
analytically and numerically.  
The $\rho^H_{_M}=2\chi \hat m^2 H^2$ 
averaged over the time $\tau_{_H}$ consistently obeys the same equation (\ref{rateeq0}) for $\rho_{_M}$,
\begin{eqnarray}
\langle \dot \rho^H_{_M} \rangle=\langle4\chi \hat m^2 H \dot H \rangle
= - \langle 2H \rho^H_{_M}\epsilon \rangle\approx -3H \rho^H_{_M},
\label{chec0}
\end{eqnarray}
where $\langle\epsilon\rangle \approx \epsilon_{_M} = 3/2$. Numerical results quantitatively show the same conclusion 
$\rho_{_M}\approx \rho^H_{_M}$,
see Fig.~\ref{balance} (a), and the detailed balance term (\ref{dbt}) vanishes, see Fig.~\ref{balance} (b). 
Thus we conclude that in the ${\mathcal M}$-{\it episode}, due to 
$\Gamma_M\gg H$ and $\tau_{_M}\ll \tau_{_H}$, the massive pairs plasma state $\rho^H_{_M}$ tightly couples with the mass density $\rho_{_M}$ in the $H$ evolution. 

\begin{figure*}[t]
\includegraphics[height=5.5cm,width=7.8cm]{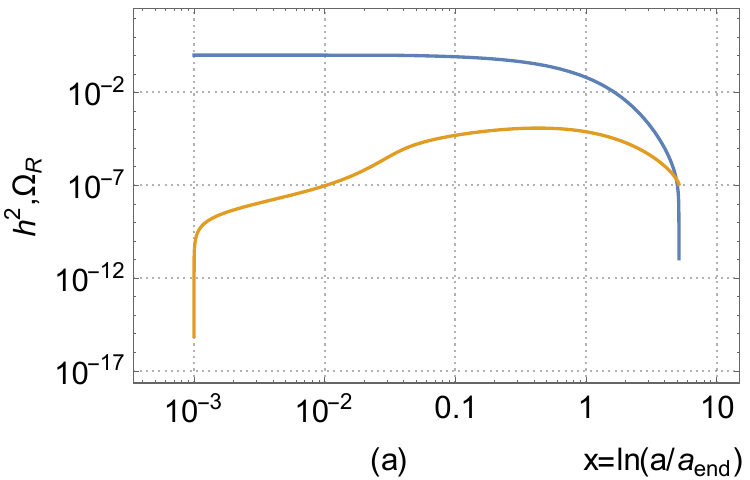}\hspace{0.333cm}
\includegraphics[height=5.5cm,width=7.8cm]{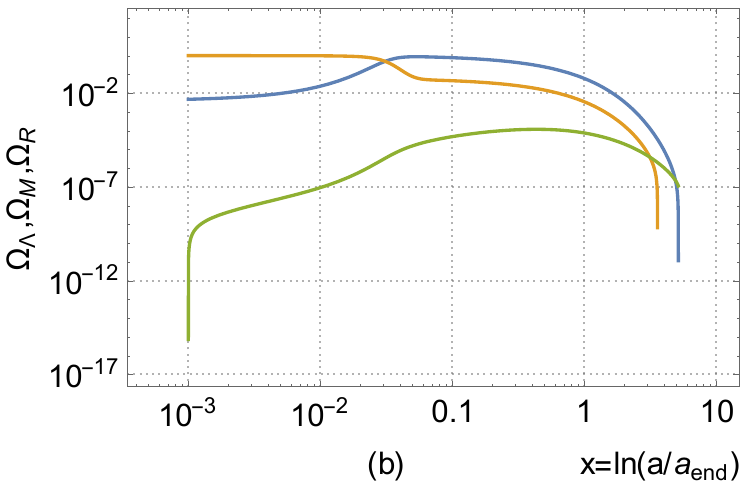}
\includegraphics[height=5.5cm,width=7.8cm]{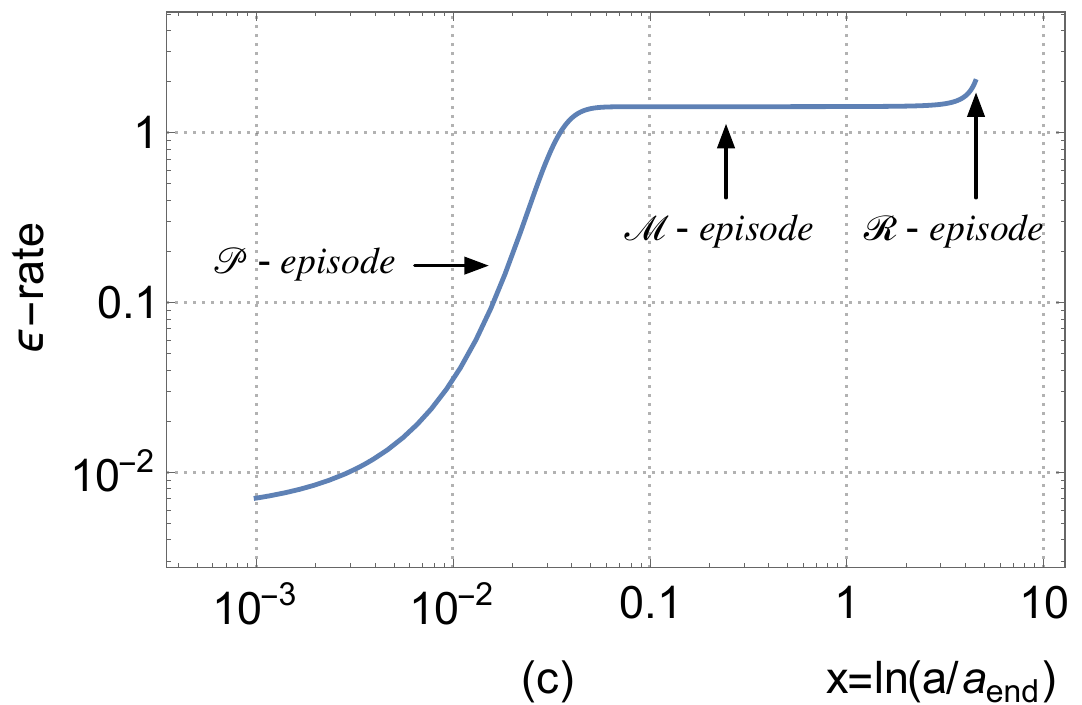}\hspace{0.333cm}
\includegraphics[height=5.5cm,width=7.8cm]{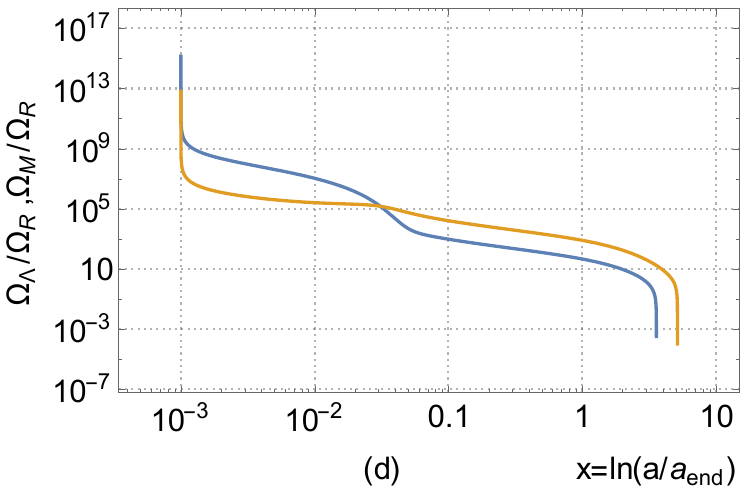}\hspace{0.333cm}
\caption{(Color Online). In a few $e$-folding number $x=\ln(a/a_{\rm end})$, 
(a) the blue line $h^2$ and orange line $\Omega_{_R}$, 
the Hubble rate drops more rapidly than the neglecting $\Omega_{_R}$ case, see 
Fig.~\ref{AllBr} (a); (b) $\Omega_{_R}$, $\Omega_{_M}$ and $\Omega_{_\Lambda}$ are 
lines green, blue, and orange; 
(c) the $H$ variation $\epsilon$-rate increases in the transition 
from $\epsilon \ll 1$ (${\mathcal P}$-{\it episode}) to the asymptotic value
$\epsilon \sim {\mathcal O}(1)$ (${\mathcal M}$-{\it episode}) and approaches to 
$\epsilon=2$ (${\mathcal R}$-{\it episode}); 
(d) the ratios of 
$\Omega_{_M}/\Omega_{_R}$ (orange) and $\Omega_{_\Lambda}/\Omega_{_R}$ (blue), recalling $h^2=\Omega_{_\Lambda}+\Omega_{_M}+\Omega_{_R}$. 
We plot these solutions with the initial conditions (\ref{ipe0}) and 
$\Omega^{\rm end}_{_R}=0.0$,
the parameter $(\hat m/m_{\rm pl})=27.7$ and 
$g^2_{_Y}=10^{-9}$.}\label{AllBrR}
\end{figure*}

\subsection{Relativistic particles domination: ${\mathcal R}$-{\it episode} 
of genuine reheating} 

At the end of the ${\mathcal M}$-{\it episode}, the massive pairs' decay term $\Gamma^{^{\rm de}}_M \rho_{_M}$ in equations 
(\ref{rateeqp},\ref{reheateq}) starts to dominate, when the time $t \gtrsim \tau_{_R}$. The $\tau_{_R}$ (\ref{Mdecayr}) is the characteristic time scale of massive pairs decay to relativistic particles. It represents the reheating period of producing tremendous amounts of entropy.  
The reheating epoch starts its genuine reheating episode, i.e.,  ${\mathcal R}$-{\it episode}.

\subsubsection{Massive and unstable pairs decay to relativistic particles} 

To study the ${\mathcal R}$-{\it episode}, we numerically solve the closed set of
the basic equations (\ref{eqhp}-\ref{reheateq}) with the radiation energy density $\Omega_{_R}$ and the decay term, 
\begin{eqnarray}
R_M=\Gamma_M^{^{\rm de}}\Omega_{_M},\quad \tau^{-1}_{_R}\equiv (R_M/\Omega_{_M})=\Gamma_M^{^{\rm de}},
\label{rmh}
\end{eqnarray}
and we define the characteristic time scale of the decay term $R_M$, which is the same as $\tau_{_R}$ (\ref{Mdecayr}). 
The initial condition of the radiation energy density is $\Omega^{\rm end}_{_R}=0$ 
at the inflation end $a_{\rm end}$, 
in addition to the initial conditions (\ref{Hend}) and (\ref{ipe0}). 
We report the numerical results in 
Figures \ref{AllBrR}. It shows that in the $H^2$ (\ref{eqhp}),
the radiation energy density 
$\Omega_{_R}$ increases and becomes dominant, 
compared with $\Omega_{_M}$ and $\Omega_{_\Lambda}$. 

\begin{figure*}[t]
\includegraphics[height=5.5cm,width=7.8cm]{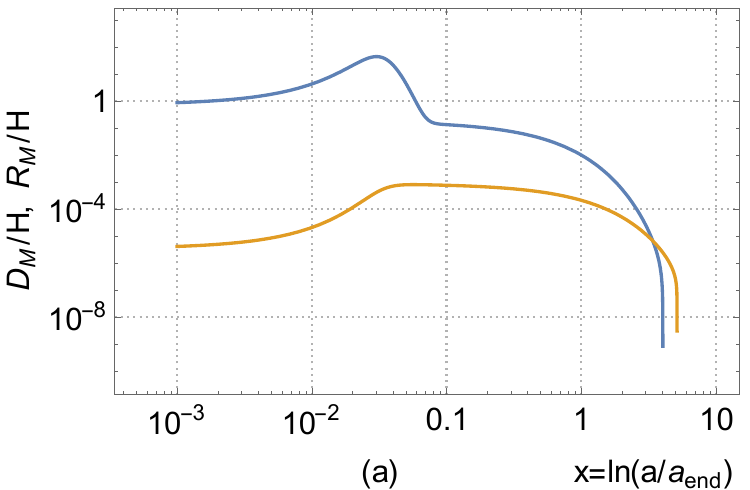}
\includegraphics[height=5.5cm,width=7.8cm]{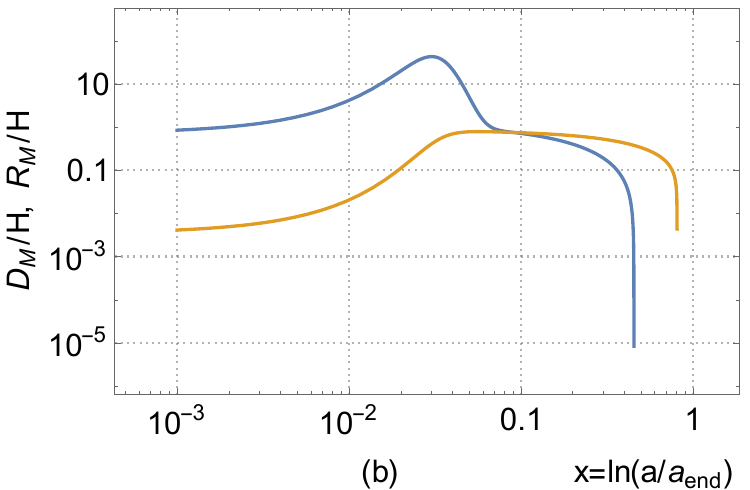} 
\includegraphics[height=5.5cm,width=7.8cm]{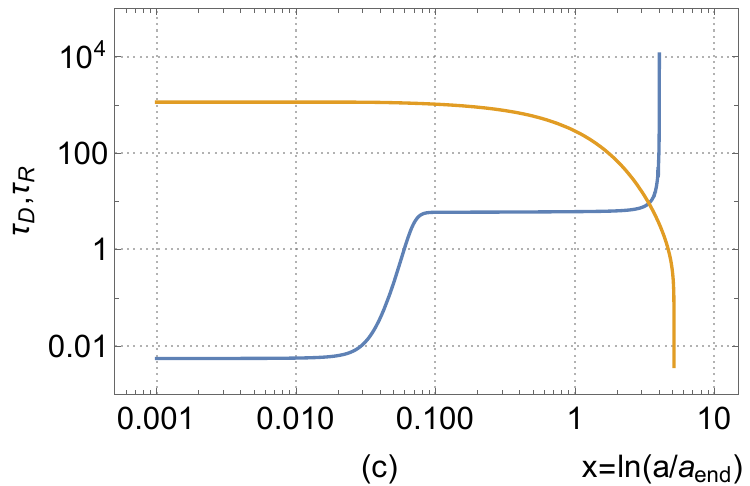}
\includegraphics[height=5.5cm,width=7.8cm]{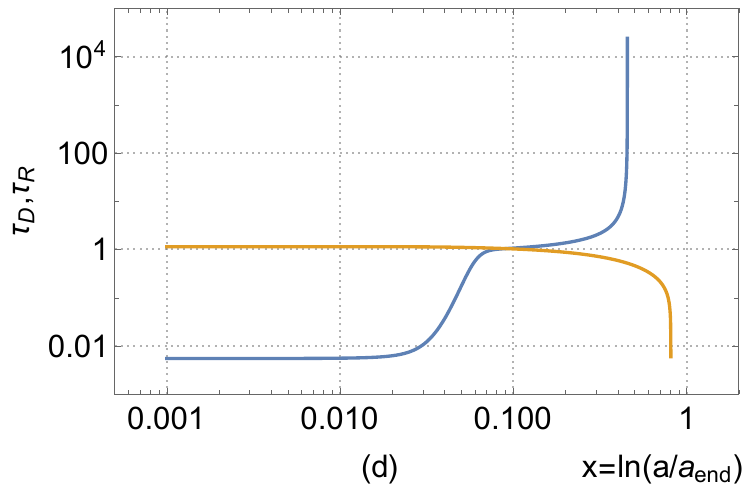} 
\caption{ (Color Online). Plotted as a function of the 
$e$-folding variable $x=\ln(a/a_{\rm end})$ for the same initial conditions 
and parameters in Figures \ref{AllBrR}, 
the detailed balance term $D_M/H$ (\ref{dbt}) (blue)
and the decay term $R_M/H$ (\ref{rmh}) (orange); two time scales $\tau_{_D}/\tau_{_H}$
(blue) and $\tau_{_R}/\tau_{_H}$ (orange). 
Left column (a) and (c) for $g^2_{_Y}=10^{-9}$ and $a_{_R}\gtrsim 20.1 a_{\rm end}$; 
Right column (b) and (d) for $g^2_{_Y}=10^{-6}$ and $a_{_R}\gtrsim 1.8 a_{\rm end}$.  
}\label{mscaleR}
\end{figure*}

We explain this phenomenon by comparing 
the decay term $R_M$ (\ref{rmh}) with the detailed balance term $D_M$ (\ref{dbt}) 
in the cosmic rate equation (\ref{rateeqp}). Two different dynamics $D_M$ and $R_M$ compete with each other in the process.  
The $\rho_{_R}$ is negligible when $D_M >R_M$,
while the $\rho_{_R}$ is dominant when $R_M>D_M$, 
and the transition from one to another occurs approximately 
at $R_M \gtrsim D_M$ , 
where $\rho_{_R}\lesssim h^2$, as shown in Figs.~\ref{mscaleR} (a) and (b). 
More precisely, it is the comparison between the characteristic time scale $\tau_{_D}$ (\ref{dbtc}) 
of the back-and-forth 
process (\ref{bfrho}) 
and the characteristic time scale $\tau_{_R}$ (\ref{rmh}) 
of the pair decay process (\ref{Mdecayr}).
When $\tau_{_D}<\tau_{_R}$, the faster back-and-forth process (\ref{bfrho}) dominates, whereas $\tau_{_R}<\tau_{_D}$, 
the faster decay process (\ref{Mdecay}) 
dominates. 

In Figs.~\ref{mscaleR} (c) and (d), two-time scales $\tau_{_D}$ and $\tau_{_R}$ are plotted as dimensionless quantities $\tau_{_D}/\tau_{_H}$ and $\tau_{_R}/\tau_{_H}$ 
to show two episodes: 
\begin{enumerate}[(i)]
\item the ${\mathcal M}$-{\it episode}, $\tau_{_D}<\tau_{_R}$ ($D_M >R_M$), 
indicating the back-and-forth process (\ref{bfrho}) dominates over the decay process (\ref{Mdecayr}); 
\item the ${\mathcal R}$-{\it episode}, $\tau_{_D}>\tau_{_R}$ ($D_M <R_M$), 
indicating the decay process (\ref{Mdecayr})  dominates over the back-and-forth process (\ref{bfrho}). 
\end{enumerate}
The separatrix of two {\it episodes} locates at $\tau_{_D}\approx \tau_{_R}$ ($D_M \approx R_M$), i.e., the crossing point of blue and orange lines in Figs.~\ref{mscaleR}.
It roughly gives the $a_{_R}$ value at which the genuine reheating occurs. 

The $a_{_R}$ value decreases as the 
Yukawa coupling $g_{_Y}$ increases, shown by the left column  (a,c) and the
right column (b,d) of Figs.~\ref{mscaleR}. 
Around this point $a_{_R}$, Figures \ref{AllBrR} (b) and (d) show 
$\Omega_{_R}\gg \Omega_{_M}\gg \Omega_{_\Lambda}$, and Fig.~\ref{AllBrR} (c) shows $\epsilon\rightarrow 2$, indicating the radiation domination.
At $a=a_{_R}$, the numerical calculations of the equations (\ref{eqhp}-\ref{reheateq}) run into the stiffness 
system of step size being effective zero. However, the analytical solution to these basic equations is studied in the next section.

\subsubsection{Energy densities of massive pairs and relativistic particles} 

The $\mathcal R$-{\it episode} is characterised by 
\begin{eqnarray}
\rho_{_R}&\gg&  \rho_{_M}\gg \rho_{_\Lambda},\quad \epsilon\rightarrow 
\epsilon_{_R}\approx 2,
\label{Mdecay}
\end{eqnarray}
and $\Gamma^{^{\rm de}}_M/H>1$, as shown in Figs.~\ref{AllBrR}. As a result, 
Equations (\ref{eqhp}) and (\ref{rehp}) or Eq.~(\ref{dde}) give 
\begin{eqnarray}
H^{-1}\approx \epsilon_{_R} t,
\quad 
a(t)/a_{_R}\approx (t/\tau_{_R})^{1/\epsilon_{_R}}.
\label{eprh}
\end{eqnarray}
Following the line presented in Ref.~\cite{Kolb1990}, we discuss how 
the massive pairs transfer their mass energy to relativistic 
particles, and calculate 
the radiation energy density $\rho_{_R}$, entropy $S$ and temperature $T$ 
of relativistic particles.

\comment{Massive pairs $\hat m$ have not much entropy, as discussed in the previous paper, at the reheating starting point, $\rho_{_M}$ is created by consuming $\rho_{_\Lambda}\approx 0$. We first assume the reheating proceeds at the initial 
condition $\rho_{_M}\gg \rho_{_\Lambda}\approx 0$ and
$\rho_{_R}$ is still very small, and thus negligible. 
$\tau_{_M} < \tau_{_R}$. 
To see how to end inflation, we need to discuss entropy and particle creation. 
The entropy has two parts: 1. particles and 2. horizon $H$ 
If the reaction rate of $\Lambda$ or space-time 
$H$ creating matter and antimatter which annihilating to $\Lambda$ is faster than Universe expansion, we assume each of them 
is in thermal equilibrium with the characteristic temperature $T_M$ or $T_H$, where Hawking temperature $T_H=H/2\pi$ of De Sitter space. 
(maybe Universe expansion is faster than some anti-matter is out of the horizon)
}

Since unstable massive pairs predominately decay to relativistic particles, the detailed balance term $D_M$ (\ref{dbt}) is negligible, and the cosmic rate equation (\ref{rateeqp}) 
reduces to, 
\begin{eqnarray}
\frac{d(a^3\rho_{_M})}{dt}&=&a^3\frac{d\rho_{_M}}{dt}+ 3Ha^3\rho_{_M} \approx - \tau^{-1}_{_R}
a^3\rho_{_M},\label{mrates0}\\
 &\Rightarrow &\rho_{_M} \approx  \rho_{_M}(a_{_R})\left(\frac{a}{a_{_R}}\right)^{-3}\exp - t/\tau_{_R}.
\label{mrates}
\end{eqnarray}
The reheating equation (\ref{reheateq}) becomes
\begin{eqnarray}
\frac{d(a^4\rho_{_R})}{dt}&=&a^3\frac{\rho_{_M}(a_{_R})}{\tau_{_R}}\left(\frac{a}{a_{_R}}\right)^{-3}\exp - t/\tau_{_R}.
\label{reh1}
\end{eqnarray} 
In theory, it requires  the 
the time integration from the initial time $t_i(a_i) \ll \tau_{_R}$ when $\rho_{_R}(a_i)=0$
to the final time $t_f\gg \tau_{_R}$ to obtain the radiation energy density
 $\rho_{_R}$,
\begin{eqnarray}
\rho_{_R}&=&\left(\frac{a_{_R}}{a}\right)^{4}\frac{\rho_{_M}(a_{_R})}{\tau_{_R}}\int^{t_f}_{t_i}
\left(\frac{t}{\tau_{_R}}\right)^{1/\epsilon_{_R}}e^{- t/\tau_{_R}}dt\approx 
0.89 \left(\frac{a_{_R}}{a}\right)^{4}\rho_{_M}(a_{_R}).
\label{reh2}
\end{eqnarray} 
Through their gauge and other induced 
interactions, these relativistic particles $\bar \ell \ell$ including sterile particles and other particles beyond the SM interact with each other. They are quickly thermalised at a very high temperature $T$, 
due to their high number and energy densities. The local thermalisation time scale 
is much shorter than the expansion time scale $\tau_{_H}=H^{-1}$, and thus the 
local thermal equilibrium is built.

\subsubsection{Reheating temperature and entropy}\label{enTem} 

We follow the approach \cite{Kolb1990} to calculate the reheating temperature and entropy. The second law of thermodynamics applied to a comoving volume element yields 
\begin{eqnarray}
dS=\frac{dQ}{T}=-\frac{d(a^3\rho_{_M})}{T}\approx  \frac{a^3\rho_{_M}}{T} \tau_{_R}^{-1} dt,
\label{1stlaw}
\end{eqnarray} 
where $dQ$ is the pair mass energy and $dS$ is the entropy of relativistic particles 
produced from massive pairs decay. Therefore, in a comoving volume, 
the entropy and energy densities of relativistic particles at the thermal state of temperature $T$ are given by,
\begin{eqnarray}
\rho_{_R}=\frac{\pi^2}{30}g_*T^4,\quad   S=\frac{2\pi^2}{45}g_* a^3T^3,\quad
\rho_{_R}=\frac{3}{4}\left(\frac{45}{2\pi^2 g_*}\right)^{1/3}S^{4/3}a^{-4}.
\label{enh}
\end{eqnarray}
The appropriately time-averaged degeneracy
$g_*$ over the decay period 
$\tau_{_R}$ counts for the total number of effectively massless degrees of freedom, 
those species share a common temperature $T$. 
Using the entropy (\ref{enh}), one writes Eq.~(\ref{1stlaw}) as,
\begin{eqnarray}
S^{1/3}\dot S = \left(\frac{2\pi^2g_*}{45}\right)^{1/3} a^4\rho_{_M} \tau^{-1}_{_R}.
\label{enheq}
\end{eqnarray}
Integrating this equation over the decay 
period $\tau_{_R}$ from the initial scale factor $a_i$ to the reheating scaling 
factor $a_{_R}>a_i$ leads to an approximate solution 
\begin{eqnarray}
S^{4/3}_R &=&1.09 \left( \frac{4}{3}\rho_{_M}(a_i)a^4_{_R}\right) \left(\frac{16\pi^3 g_*\rho_{_M}(a_i) }{135 M^2_{\rm pl}}\right)^{1/3}\tau^{2/3}_{_R},\nonumber\\
\Rightarrow S_R &\approx &1.32(16\pi^3g_*/135)^{1/4}a^3_{_R} \rho_{_M}(a_i) (\tau_{_R}/M_{\rm pl})^{1/2}.
\label{senheq}
\end{eqnarray}
Here one adopts Eq.~(\ref{mrates}) and  
$S_i(a_i)\approx 0$, namely, the initial massive pairs' entropy is approximately zero. 
In principle, it requires integrating from the initial 
time $t_i(a_i) \ll \tau_{_R}$ to the final time $t_f(a_{_R})\gg \tau_{_R}$, when the 
entropy significantly increases. In practice, 
$a_i\lesssim a_{_R}$ and $t_f\gtrsim \tau_{_R}$ are approximately 
adopted in Eq.~(\ref{senheq}), since the all-important entropy $S_R$ mainly produces in the reheating period $\tau_{_R}(a_{_R})$.

At the scale factor $a_{_R}$,
the reheating scale $H_{\rm RH}$ can be obtained by 
$\tau_{_R}$ from the Friedmann equation (\ref{eqhp}) 
or the reheating temperature $T_{\rm RH}\equiv T(t=\tau_{_R})$ from 
the thermalization (\ref{enh}) \cite{Kolb1990}:
\begin{eqnarray}
H^2_{\rm RH}&\equiv& H^2(t=\tau_{_R})\approx \frac{1}{4}\tau^{-2}_{_R},\label{ent0}\\
H^2_{\rm RH}&\approx &\frac{8\pi}{3M^2_{\rm pl}}\rho_{_R}\approx\frac{8\pi}{3M^2_{\rm pl}}\left(\frac{\pi^2g_*}{30}T^4_{\rm RH} \right).\label{ent1}
\end{eqnarray}
It leads to the reheating temperature 
\begin{eqnarray}
T_{\rm RH}\approx 0.55 g_*^{-1/4}(M_{\rm pl}/\tau_{_R})^{1/2}
=0.55 (g_{_Y}^2/g^{1/2}_*)^{1/2}(\hat m /M_{\rm pl})^{1/2}M_{\rm pl},
\label{ent}
\end{eqnarray} 
and the all-important entropy per comoving volume,
\begin{eqnarray}
S_R &\approx &1.32\left(\frac{16\pi^3}{135}\right)^{1/4} (g_*^{1/2}/g^2_{_Y})^{1/2} 
(\hat m/M_{\rm pl})^{1/2}a^3_{_R} \rho_{_M}(a_i)/\hat m.
\label{senheq1}
\end{eqnarray}
Equations (\ref{ent0}) and (\ref{ent1}) physically 
mean that at the the genuine reheating (i) $H_{\rm RH}\approx \Gamma^{^{\rm de}}_M/2=g_{_Y}^2\hat m/2$
the Hubble rate is in the same order as the pair decay rate; (ii) 
$H^2_{\rm RH}\approx \rho_{_R}/3m^2_{\rm pl}$ the radiation energy is 
predominate.   
These results depend on the effective degeneracy $g_*$ (\ref{enh}) and the decay rate
$\tau^{-1}_{_R}=g_{_Y}^2\hat m$ (\ref{Mdecayr}). 

Our numerical calculations show the consistency of 
the approximation $a_i\lesssim a_{_R}$ 
used in Eq.~(\ref{senheq}) and the agreement with 
the analytical solutions (\ref{ent0}) and (\ref{ent}). 
From Figs.~\ref{AllBrR} and \ref{mscaleR}, 
we find that the reheating predominately takes place around $a_{_R}$, 
at which $\tau_{_R}\sim \tau_{_D}$, and the ratios $\tau_{_R}/H_{\rm RH}\approx \tau_{_D}/H_{\rm RH} \sim {\mathcal O}(1)$.
Moreover, from Fig.~\ref{AllBrR} ($a$) we obtain 
the reheating scale 
\begin{eqnarray}
H_{\rm RH}\approx 3.16\times 10^{-4}H_{\rm end}\approx 6.1\times 10^{9}{\rm GeV},\quad a_{_R}\approx 20.1 a_{\rm end}
\label{hreh}
\end{eqnarray}
for the case $g^2_{_Y}=10^{-9}$ and $\hat m=27.7m_{\rm pl}$. We also obtain  
$H_{\rm RH}\sim 10^{-1}H_{\rm end}= 1.9\times 10^{12}{\rm GeV}$  (the plot is not present), 
and $a_{_R}\approx 1.8 ~a_{\rm end}$ for the case $g^2_{_Y}=10^{-6}$. 

\comment{
\subsubsection{Reheating scale factor and initial condition 
for standard cosmology}\label{icsc}
Our study shows that the reheating epoch composes the preheating 
${\mathcal P}$-episode, the ${\mathcal  M}$-episode of massive pairs domination, 
and the genuine reheating ${\mathcal  R}$-episode, and
it lasts for the period from $a_{\rm end}=a_3$ to $a_{_R}=a_2$, 
see Fig.~\ref{schematicf}.  
From the numerical results 
$\epsilon$-rate of Figs.~\ref{AllBrR} (c) and 
$\rho_{_M}\gg \rho_{_\Lambda}$ of Fig.~\ref{mscale} (right) for the mass parameter 
$\hat m > 20m_{\rm pl}$ (\ref{mthre}), we observe that the ${\mathcal  M}$-episode
$\epsilon\approx 3/2$ lasts last much longer time than 
${\mathcal P}$-episode $\epsilon\ll 1$ and the ${\mathcal  R}$-episode 
$\epsilon\approx 2$. This implies that in the reheating epoch the 
$\rho_{_\Lambda}\approx 3m^2_{\rm pl}H^2_{\rm end}$
the energy density of the spacetime converts to 
the $\rho_{_M}$ energy density of massive pairs, which then converts to the 
$\rho_{_R}\approx 3m^2_{\rm pl}H^2_{\rm RH}$ energy density 
of relativistic particles at the genuine reheating (\ref{1stlaw}). 
This is the case that the reheating epoch ends at the scale factor $a_{_R}$
with the condition $\rho_{_\Lambda}\ll \rho_{_M}\ll \rho_{_R}$ to 
initiate the standard cosmological scenario}

To estimate the the scale factor change $a_{_R}/a_{\rm end}$ 
in the reheating epoch, 
we approximately use the conservation law (\ref{rateeq0}) for the 
massive pair domination
\begin{eqnarray}
\Delta_2\equiv\frac{a_{_R}}{a_{\rm end}}\approx 
\left(\frac{\rho^i_{_M}}{\rho^f_{_M}}\right)^{1/3}\approx \frac{1}{\pi}
\left(\frac{45}{4}\right)^{1/3}\left(\frac{H^2_{\rm end}M_{\rm pl}^2}{g_*T^4_{\rm RH}}\right)^{1/3}.
\label{d2}
\end{eqnarray}
Here the initial pair energy density 
$\rho^i_{_M}\approx\rho^{\rm end}_{_\Lambda}\approx 3m^2_{\rm pl}H^2_{\rm end}$
(\ref{ratioend}) at the beginning of the reheating epoch, and 
the final one $\rho^f_{_M}\approx \rho_{_R}\approx 3m^2_{\rm pl}H^2_{\rm RH}$ 
at the end of the reheating epoch, in virtue of Eqs.~(\ref{enh}) and (\ref{ent1}). 

\comment{
As a result, $\Omega_{_\Lambda}$ almost decouples from 
Eq.~(\ref{reh}), as if it had been frozen as a ``constant''. Nevertheless $\Omega_{_\Lambda}$ weakly coupling to $\Omega^{\rm l,h}_{_M}$ via $h$ and dominantly governs $h$ again, as decreasing 
$\Omega^{\rm l,h}_{_M}< \Omega_{_\Lambda}$. 
because $\Gamma \ll H$ and $T_M \gg T_H$ in this epoch, particles 
have not enough density and rate to annihilate back to the spacetime.
(iii) $\Gamma < H$, $T_M > T_H$. The spacetime decouples from matter particles, the latter evolves with this huge entropy in standard cosmology, however, there is still interaction between spacetime and matter particles, though they are decoupled from each other from thermal equilibrium, and not in equilibrium. The particle matter density value at the heating era is given by Eq.~(\ref{nden}) at $H_{\rm end}$, 
and then follows the law $(a_{\rm end}/a)^4$ to decrease, assume they are a relativistic gas. This plays relevant roles in reheating, baryogenesis and magnetogenesis. How about observations to do with the ratio $\rho_{_\Lambda}/\rho_{_M}$, where $\rho_{_M}$ is converted to $\rho_{_R}$.
}

\section{Observations to fix reheating temperature and entropy}\label{obsm} 

We use the method proposed by Ref.~\cite{Mielczarek2011} 
to fix the reheating temperature by the CMB observations.
The cosmological evolution of the physical wavelength $\lambda(a)$ and wavenumber $k(a)$ is
\begin{eqnarray}
\lambda(a)=\lambda_0\frac{a}{a_0},\quad k(a)=k_0\frac{a_0}{a}\quad \lambda(a)=\frac{1}{k(a)},
\label{wavel}
\end{eqnarray}
where the present time $a_0=1$, the comoving wavenumber $k(a_0)$ and wavelength $\lambda_0=1/k_0$ are constants in the evolution, see Figure \ref{schematicf}.
The total increase of the scale factor from the horizon crossing $a_*$ (\ref{Hend}) to $a_0$ is given by 
\begin{eqnarray}
\Delta_{\rm tot}=\frac{a_0}{a_*}=\frac{\lambda_0}{\lambda(a_*)}=\frac{H_*}{k_0}.
\label{totala}
\end{eqnarray}
At the CMB pivot scale $\lambda(a_*)=H^{-1}_*=k^{-1}_*$, the scalar spectrum gives 
\begin{eqnarray}
\Delta_{\rm tot}=\frac{M_{\rm pl}}{k_*}\sqrt{\pi\epsilon_*A_s}=\frac{M_{\rm pl}}{\sqrt{2}~k_*}\sqrt{\pi(1-n_s)A_s}.
\label{totalav}
\end{eqnarray}
On the other hand, as illustrated in Fig.~\ref{schematicf} and Ref.~\cite{Mielczarek2011},
$\Delta_{\rm tot}=\Delta_3\Delta_2\Delta_1\Delta_0$,  
$\Delta_1=(a_{\rm rec}/a_{_R})=(g_*/2)^{1/3}(T_{\rm RH}/T_{\rm rec})$ 
and $\Delta_0=(a_0/a_{\rm rec}) = 1+ z_{\rm rec}$ are given in terms of the temperature $T_{\rm rec}=T_{\rm CMB}(1+ z_{\rm rec})$ and redshift $z_{\rm rec}$ at the recombination, 
\begin{eqnarray}
\Delta_1\Delta_0=\frac{a_0}{a_{_R}}\approx (g_*/2)^{1/3}(T_{\rm RH}/T_{\rm CMB}).
\label{d10}
\end{eqnarray}
Whereas, we compute $\Delta_3=(a_{\rm end}/a_*)$ 
(\ref{Hend}) and $\Delta_2=(a_{_R}/a_{\rm end})$ (\ref{d2}) in the $\tilde\Lambda$CDM scenario. 
As a result, we obtain
\begin{eqnarray}
\Delta_{\rm tot}=e^{N_{\rm end}}\frac{1}{\pi}
\left(\frac{45}{4}\right)^{1/3}\left(\frac{H^2_{\rm end}M_{\rm pl}^2}{g_*T^4_{\rm RH}}\right)^{1/3}\frac{T_{\rm RH}}{T_{\rm CMB}}\left(\frac{g_*}{2}\right)^{1/3}.
\label{totalav1}
\end{eqnarray}
It agrees with the result (33) using the inflation potential $V(\phi)$ energy density (17) in 
Ref.~\cite{Mielczarek2011}. Here we adopt the energy density $\rho^{\rm end}_c\equiv 3
m^2_{\rm pl}H^2_{\rm end}$ (\ref{ratioend}) at inflation end.  

Equations (\ref{totalav}) and (\ref{totalav1}) are independent of the effective
reheating degeneracy $g_*$ and yield the reheating temperature
\begin{eqnarray}
\frac{T_{\rm RH}}{M_{\rm pl}}&=&\left(\frac{45}{2^{3/2}}\right)\frac{e^{3N_{\rm end}}}{\pi^{9/2}}
[(1-n_s)A_s]^{-3/2}\left(\frac{k_*}{T_{\rm CMB}}\right)^3 \left(\frac{H_{\rm end}}{M_{\rm pl}}\right)^2\label{rehtem},
\end{eqnarray}
in terms of the CMB observations $T_{\rm CMB}
=2.725 ~{\rm K} = 2.348\times 10^{-4}$ eV and $k_*= 0.05{\rm Mpc}^{-1}$ (
${\rm Mpc}^{-1}=6.39\times 10^{-30}$eV), as well as $N_{\rm end}$ (\ref{endr})
and $H_{\rm end}$ (\ref{Hend}), 
whose values depend on the CMB measurements 
$A_s$, $n_s$ 
and $r$, see Sec.~\ref{infend}. 
 
\begin{figure*}[t]
\includegraphics[height=5.5cm,width=7.8cm]{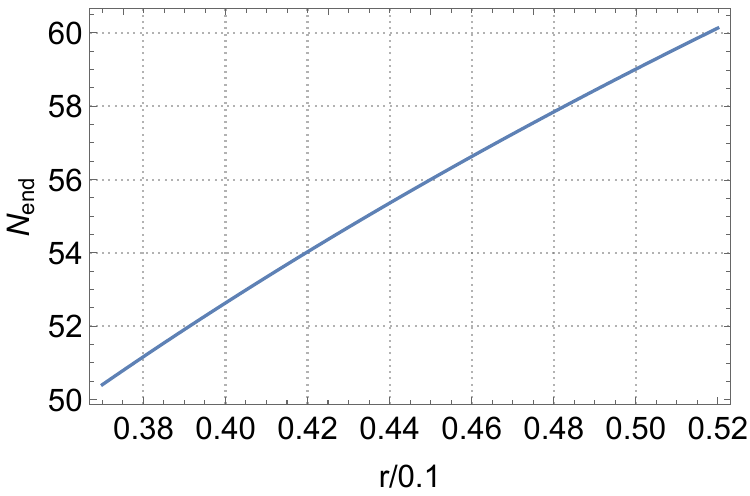}
\includegraphics[height=5.5cm,width=7.8cm]{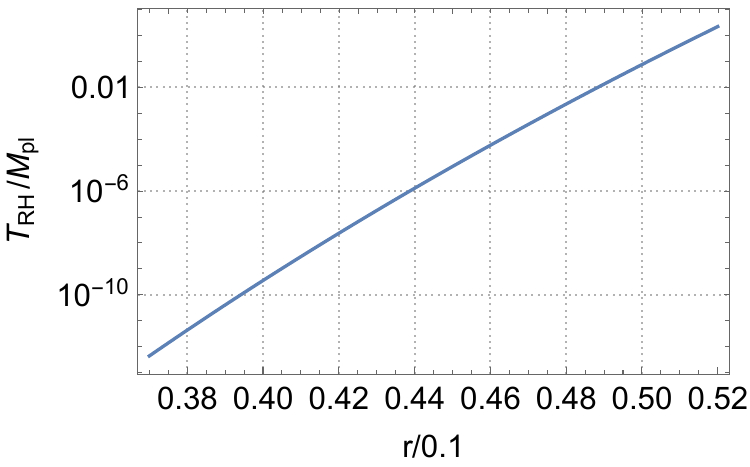}
\caption{ (Color Online).  
Fixed the observed spectral index $n_s=0.965$, 
in terms of the tensor-to-scalar ratio $r$, we plot 
the inflation end $e$-folding number 
$N_{\rm end}$ (\ref{endr}) and the reheating temperature $T_{\rm RH}$ (\ref{rehtem}). 
These plots refer to their lower limits due to the nature of inequality (\ref{endr}). 
The real values of $N_{\rm end}$ and $T_{\rm RH}$ should be slightly above 
the curves for a given $r$ value.  
}\label{Treheat}
\end{figure*}

\subsection{Reheating temperature and entropy vs tensor-to-scalar ratio 
$r<0.048$}   

Given the observed the scalar amplitude $A_s=2.1\times 10^{-9}$ and 
spectral index $n_s=0.965$,  
the inflation ending $e$-folding number 
$N_{\rm end}$ (\ref{Hend}) and the reheating temperature $T_{\rm RH}$
(\ref{rehtem}) are plotted in Fig.~\ref{Treheat} as functions of the 
tensor-to-scalar ratio $r$ without any free parameter. 
Figure \ref{Treheat} shows that their  values are
$N_{\rm end}\approx (50,60)$ and 
$T_{\rm RH}/M_{\rm pl}\approx (5.5\times 10^{-13},1.1\times 10^{0})$ in the range 
$r \approx (0.037,0.052)$. 
 
\begin{figure*}[b]
\includegraphics[height=5.5cm,width=7.8cm]{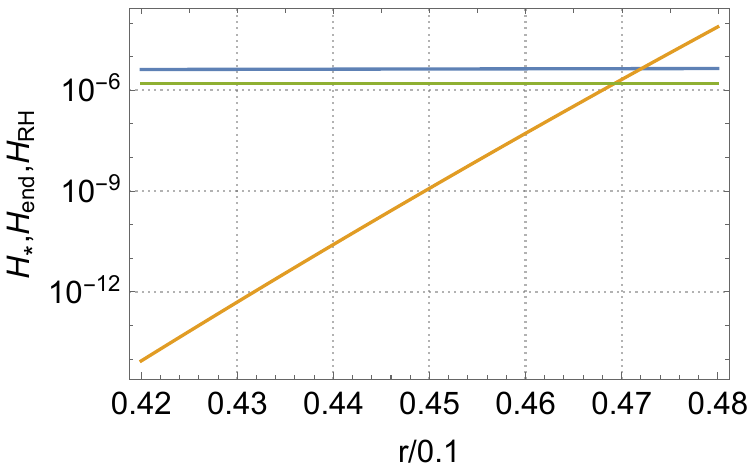}
\includegraphics[height=5.5cm,width=7.8cm]{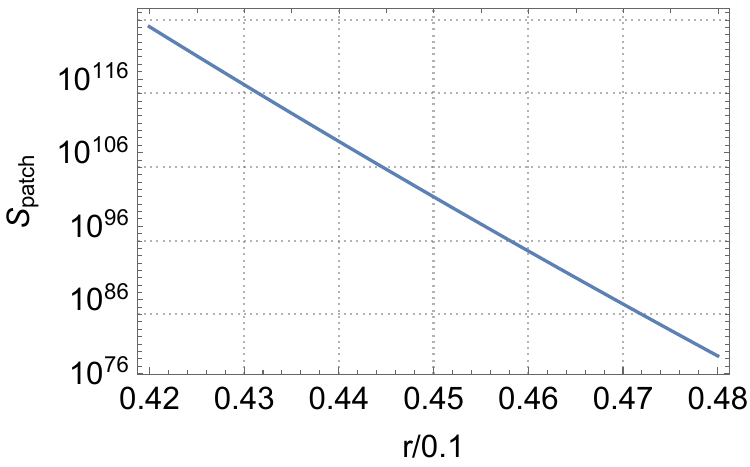}
\caption{ (Color Online).  
Fixed the observed spectral index $n_s=0.965$, as functions
of the tensor-to-scalar ratio $r$, we plot (Left)
the inflation scale $H_*$ (blue), inflation end scale 
$H_{\rm end}$ (\ref{Hend}) (green), and reheating scale 
$H_{\rm RH}$ (\ref{ent1}) (orange) in unit of the Planck scale $M_{\rm pl}$;
(Right) the entropy $S_{\rm patch}$ (\ref{senheq3}) 
within the physical patch $H^{-3}_{\rm RH}$ 
at the reheating end $a_{_R}$. $H_{\rm RH}\propto g_*^{1/2}$ and $S_{\rm patch}\propto g_*^{-5/2}$. 
We adopt $g_*\simeq 10^2$ for the standard model of particle physics, including sterile neutrinos.
}\label{entropy-vs-rf}
\end{figure*}

After obtaining the reheating temperature at the end $a_{_R}$ of reheating, 
we calculate the entropy $S_{\rm patch}$ produced within the physical patch of 
the volume $H^{-3}_{\rm RH}$, which evolves from the initial patch 
of the volume $H^{-3}_*$ at the start $a_*=1$ 
of the inflation. 
The patch grows by a scale factor of Eqs.~(\ref{Hend}) and (\ref{d2}),
\begin{eqnarray}
a^3_{_R}=(\Delta_3\Delta_2)^3\approx  
\frac{45 e^{3N_{\rm end}}}{4\pi^3}
\left(\frac{H^2_{\rm end}M_{\rm pl}^2}{g_*T^4_{\rm RH}}\right)=e^{3N_{\rm end}}
\left(\frac{H^2_{\rm end}}{H^2_{\rm RH}}\right).
\label{d32}
\end{eqnarray}  
The entropy per comoving volume $S_R$ (\ref{senheq1}) at the end of reheating 
is expressed as
\begin{eqnarray}
S_R &\approx &2.2\times 10^{6}\left(\frac{16\pi^3}{135}\right)^{1/4}
\frac{3a^3_{_R}}{8\pi} H^2_{\rm end}M_{\rm pl}, 
\label{senheq2}
\end{eqnarray}
where we use $\rho^i_{_M}(a_i)\approx\rho^{\rm end}_{_\Lambda}\approx 3m^2_{\rm pl}H^2_{\rm end}$ (\ref{d2}) and the constraint (\ref{para1}) below. 
The entropy $S_{\rm patch}$ within the physical patch $H^{-3}_{\rm RH}$ is given by,
\begin{eqnarray}
S_{\rm patch} =  H^{-3}_{\rm RH} S_R &\approx & 3.64\times 10^{5} ~e^{3N_{\rm end}} \left(\frac{H^4_{\rm end}M_{\rm pl}}{H^5_{\rm RH}}\right),
\label{senheq3}
\end{eqnarray}
which is a function of $r$ and $g_*$. 
In Fig.~\ref{entropy-vs-rf} (right),  we plot $S_{\rm patch}$ 
by using $N_{\rm end}$ (\ref{endr}), $H_{\rm end}$ (\ref{Hend}) 
and $H_{\rm RH}$ (\ref{ent1}). It shows the calculated entropy accords with 
the observational value $S_{\rm patch}\sim 10^{88}$ around $r \sim 0.045$. 

To have a better understanding of how the physical patch horizon 
$H_*> H_{\rm end}> H_{\rm RH}$ evolves,
we plot in the same Fig.~\ref{entropy-vs-rf} (left) all characteristic 
Hubble scales from the inflation to the reheating:
the inflation scale $H_*$, inflation end scale 
$H_{\rm end}$ (\ref{Hend}), and reheating scale 
$H_{\rm RH}$ (\ref{ent1}) in unit of the Planck scale $M_{\rm pl}$. It shows that
the nonphysical situation $H_{\rm RH}>H_{\rm end}$ occurs when $r>0.047$. 
Therefore the $r>0.047$ range excludes. Due to 
the dependence of $H_{\rm RH}$ on $g_*$ and the approximations adopted in these calculations, we conservatively suggest a theoretical 
upper limit of the tensor-to-scalar ratio $r < 0.047$.
This theoretical upper limit 
is consistent with the observational one $r<0.065$ \cite{Akrami2020}, and  
the recent constraint $r<0.044$ \cite{Tristram2021}.

In the $r$-range $[0.042, 0.048]$, we show that (i)
the inflation $e$-folding number $58\gtrsim N_{\rm end}\gtrsim 54$ and the reheating temperature 
$10^{-3}\gtrsim T_{\rm RH}/M_{\rm pl}\gtrsim  10^{-8}$ from the numerical results
presented in Fig.~\ref{Treheat}; (ii) 
the inflation scale 
$H_*/M_{\rm pl}\approx 4.0\times 10^{-6}$, 
inflation end scale $H_{\rm end }/M_{\rm pl}\approx 1.5\times 10^{-6}$, 
whereas $10^{-14}\gtrsim H_{\rm RH}/M_{\rm pl}\gtrsim  10^{-4}$ and the entropy 
$10^{120}\gtrsim S_{\rm patch}\gtrsim  10^{76}$ from the numerical results
presented in Fig.~\ref{entropy-vs-rf}.
These results show that the $\tilde\Lambda$CDM scenario is consistent with observations. The 
precisely measuring $r$-value is essential 
to determine the 
$e$-folding number of the inflation, the reheating temperature, 
all characteristic Hubble scales and produced entropy. 

\subsection{Genuine reheating $\rho_{_R}\gg \rho_{_M} \gg \rho_{_\Lambda}$ and tensor-to-scalar ratio $r>0$}\label{fixpara}

In the $\tilde\Lambda$CDM scenario, there are two parameters to describe the properties 
of the reheating epoch: (i) the effective mass parameter $\hat m/M_{\rm pl}$ physically 
represents pairs' masses and numbers;
(ii) the effective Yukawa coupling $(g_{_Y}^2/g^{1/2}_*)$ represents pairs' decay 
strength to relativistic particles of degeneracies $g_*$. 

To determine these two parameters,
we use Eqs.~(\ref{ent}) and (\ref{rehtem}) to obtain one constraint,
\begin{eqnarray}
\left(\frac{T_{\rm RH}}{M_{\rm pl}}\right)^2=0.3~(g_{_Y}^2/g^{1/2}_*)~(\hat m/M_{\rm pl}).
\label{para0}
\end{eqnarray}
Another constraint is 
\begin{eqnarray}
(g_{_Y}^2/g^{1/2}_*)~(M_{\rm pl}/\hat m)\approx 3.6\times 10^{-13},
\label{para1}
\end{eqnarray}
from the reheating temperature $T_{\rm RH}$ (\ref{ent}) 
and the ratio $T_{\rm RH}/\hat m\approx 3.3\times 10^{-7}$, obtained by the baryon number-to-entropy ratio  $n_{_B}/S_R=  0.864^{+0.016}_{-0.015}\times 10^{-10}$ \cite{Ade2016},
theoretical relation $n_{_B}/S_R\approx \epsilon_{CP}(T_{\rm RH}/\hat m)$ \cite{Kolb1990} and $\epsilon_{CP}\approx 2.6\times 10^{-4}$, see Eq.~(5.3) of Ref.~\cite{Xue2020b}.

\begin{figure*}[t]
\includegraphics[height=5.5cm,width=7.8cm]{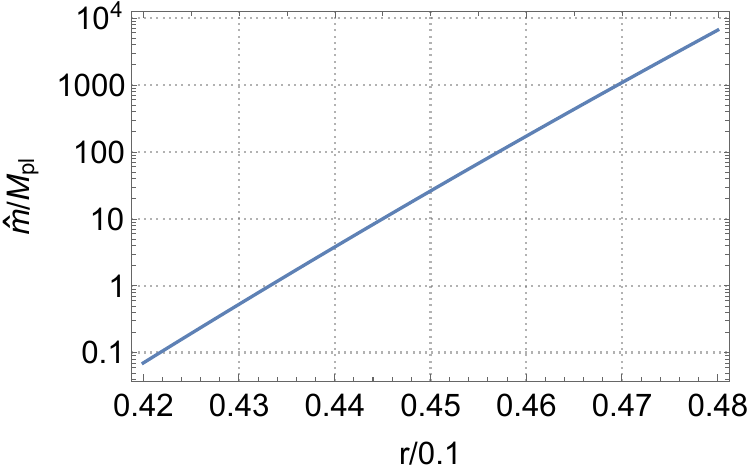}
\includegraphics[height=5.5cm,width=7.8cm]{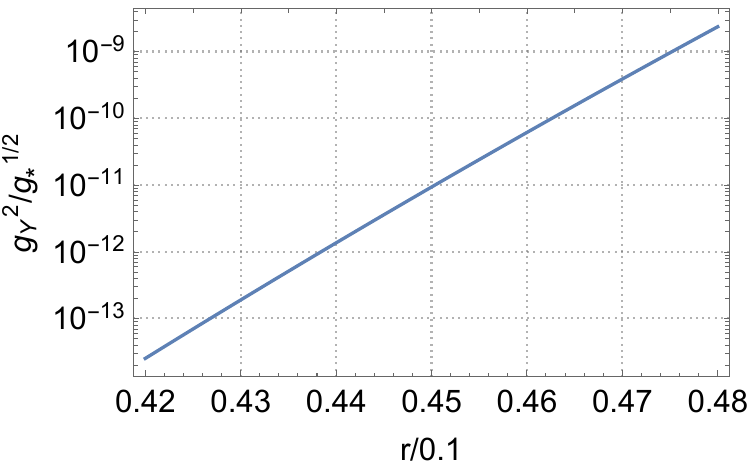}
\caption{ (Color Online).  
Using the observed spectral index $n_s=0.965$, we plot 
the mass parameter $\hat m/ M_{\rm pl}$ (left) and 
the effective Yukawa coupling $(g_{_Y}^2/g^{1/2}_*)$ (right), as functions
of the tensor-to-scalar ratio $r$ in the same range $(0.042,0.048)$ where the 
physically sensible values $N_{\rm end}$ and $T_{\rm RH}$ 
are also plotted in Fig.~\ref{Treheat}.
}\label{mscale-vs-rf}
\end{figure*}

In Fig.~\ref{mscale-vs-rf}, we numerically plot
the constraints (\ref{para0}) and (\ref{para1}) 
as a function of the tensor-to-scalar ratio $r$. 
Recall that in Sec.~\ref{threshold} 
we point out the theoretical threshold 
$\hat m_{\rm thresh}$: $\hat m> 20 m_{\rm pl} = 4 M_{\rm pl}$ (\ref{mthre}) for $\rho_{_M}\gg \rho_{_\Lambda}$. We conclude that the nontrivial threshold $\hat m_{\rm thresh}$ demands the tensor-to-scalar 
ratio $r>0$, see also 
Eq.~(\ref{endr}). Applying the obtained theoretical threshold $\hat m_{\rm thresh}=4 M_{\rm pl}$ 
to numerical results 
Fig.~\ref{mscale-vs-rf}, we find that the tensor-to-scalar 
ratio $r\gtrsim 0.044$. However, it depends on the parameter $\chi\approx 1.85\times 10^{-3}$ adopted and other approximations used for 
numerical calculations. To obtain the lower limit $r_{\rm min}\not=0$, it requires 
more elaborated calculations, for example, separating in Eq.~(\ref{apdenm}) unstable modes' contribution 
$\rho^H_{_M}|_{\rm unstable}$ from stable modes' contribution $\rho^H_{_M}|_{\rm stable}$ to more accurately determine the mass parameter threshold $\hat m_{\rm thresh}$.
Nevertheless, the obtained $r$-range ($0<r<0.047$) is 
relevant to the measurements by the next generation CMB observations, such as CMB-S4 
which measures $r\gtrsim 10^{-3}$ \cite{Abazajian2019}. 

We report the numerical results in the $r$ range $[0.042, 0.048]$. 
The effective pair mass parameter is in
$10^{-1}\lesssim \hat m/M_{\rm pl}\lesssim 10^{4}$ 
and the effective Yukawa coupling is in
$10^{-14}\lesssim  (g_{_Y}^2/g^{1/2}_*) \lesssim 10^{-9}$.  
If $g_*\gtrsim 10^2$ for the standard model of particle physics, including sterile neutrinos, the Yukawa coupling is in the range of
$10^{-13}\lesssim g_{_Y}^2 \lesssim 10^{-8}$. 
We check back the parameters used in Figs.~\ref{AllBrR} and \ref{mscaleR}
are $(\hat m/m_{\rm pl})=27.7$ and $g^2_{_Y}=10^{-9}$. 
It indicates that the $\tilde\Lambda$CDM scenario 
is self-consistent and self-contained.

\section{Stable massive pairs and cold dark matter abundance}\label{coldm}

To further clarify dark energy and matter interaction, we study stable massive pairs after reheating. 
These pairs neither decay into light particles nor contribute to reheating. Thus, they remain and can be candidates for massive cold dark matter (CDM) particles. Here, we qualitatively discuss their evolution after reheating.

In the radiation- and matter-dominated epoch after reheating, the Universe evolution $\epsilon$ rate (\ref{dde0}) is $\epsilon \approx 2$ and $\epsilon \approx 3/2$, respectively. The mean pair-production rate (\ref{prate}) $\Gamma_M\propto m$ is proportional to the mass parameter $m$. The large mass $m$ and ratio $\Gamma_M/H>1$ imply massive stable pairs should tightly couple with dark energy in evolution. Therefore, we focus on stable pairs interacting with dark energy in 
Eqs.~(\ref{rhomm},\ref{rholm}),  and approximately obtain  
\begin{eqnarray}
\frac{d\rho_{_\Lambda}}{dx} &\approx & - ( \Gamma_M/H)
(\rho^H_{_M}|_{\rm stable}-\rho^{\rm cold}_{_M}),\label{rhommc}\\
\frac{d\rho^{\rm cold}_{_M}}{dx} + 3 \rho^{\rm cold}_{_M} &\approx &  + ( \Gamma_M/H)(\rho^H_{_M}|_{\rm stable}-\rho^{\rm cold}_{_M}),
\label{rholmc}
\end{eqnarray}
where $\rho^H_{_M}|_{\rm stable}$ indicates the massive pair plasma state (\ref{apdenm}) for stable pairs. 
For the coupled case $\Gamma_M/H>1$, 
discussed as the case (iii) after Eqs.~(\ref{rhomm},\ref{rholm}), the approximate solution to Eqs.~(\ref{rhommc},\ref{rholmc}) is
\begin{eqnarray}
\rho^{\rm cold}_{_M}\approx \rho_{_M}^H|_{\rm stable}=2\chi (\hat m_{_M})^2|_{\rm stable}H^2.
\label{cdmb0}
\end{eqnarray}
Equations (\ref{rhommc},\ref{rholmc}) become
\begin{eqnarray}
\dot \rho_{_\Lambda} \approx  0,\quad
\dot \rho^{\rm cold}_{_M} + 3 H\rho^{\rm cold}_{_M} \approx   -\dot \rho_{_\Lambda},
\end{eqnarray}
with approximate solutions $\rho^{\rm cold}_{_M}\propto (1/a)^{(3-\delta_{_M})}$ and $\rho_{_\Lambda}\propto (1/a)^{\delta_{_\Lambda}}$, where indexes $|\delta_{_{M,\Lambda}}|\ll 1$ \cite{Xue2015}. It shows massive cold dark matter approximately follows the evolution of non-relativistic fluid, weakly interacting with dark energy. In other words, via massive pair plasma state $\rho_{_M}^H|_{\rm stable}$, dark energy and cold dark matter interact with each other. 
However, the exchange between them is small and inefficient. Actually, Eq.~(\ref{cdmb0}) is essentially the same as Eq.~(\ref{avede}) 
and discussions here are similar to the matter-dominated ${\mathcal M}$-episode of reheating, see Sec.~\ref{Msec}.  One can also find similar 
discussions \cite{Xue2022} that dark energy weakly interacts with radiation and baryon matter, and their exchange is small and inefficient.

\comment{For self consistency of solution (\ref{cdmb0}), we check if $\rho_{_M}^H$ approximately follows the continuity equation $\dot\rho^H_{_M} + 3H \rho^H_{_M} \approx 0$, indeed, 
\begin{eqnarray}
\dot \rho^H_{_M}|_{\rm stable}= 4\chi \hat m^2|_{\rm stable} H \dot H
= - 2H \rho^H_{_M}|_{\rm stable}\epsilon \approx -3H \rho^H_{_M},
\label{chec0}
\end{eqnarray}
where the Universe evolution rate (\ref{dde0}) $\epsilon \approx  3/2$ in the matter dominated epoch and $\epsilon \approx  2$ in the radiation-dominated epoch. In both cases, the continuity equation holds approximately, but why. }

Equation (\ref{cdmb0}) 
indicates the cold dark matter abundance $\Omega^{\rm cold}_{_M}$ 
is an approximate constant in time, 
\begin{eqnarray}
\Omega^{\rm cold}_{_M}\approx \Omega_{_M}^{H}\big|_{\rm stable}=\frac{\rho^H_{_M}|_{\rm stable}}{\rho^H_c}=\frac{16\pi}{3}\chi\left(\frac{\hat m_{_M}|_{\rm stable}}{M_{\rm pl}}\right)^2,
\label{cpara3}
\end{eqnarray}
where $\rho^H_c\equiv 3H^2/(8\pi G)$. It means that (i) the CDM composes stable massive pairs produced in the pre-reheating ${\mathcal P}$-episode, see Sec.~\ref{pre}; (ii) the current CDM abundance 
is approximately equal to its abundance in the ${\mathcal M}$-episode (\ref{Msec}) of reheating, see Sec.~\ref{Msec}. The approximate constancy of CDM abundance in time is a direct consequence of the holographic and massive pair plasma state (\ref{apdenm}).
To be self-consistent, we give a rough check on the $\Omega^{\rm cold}_{_M}$ magnitude by using 
the theoretical threshold (\ref{mthre}) $\hat m_{_M}|_{\rm stable}/M_{\rm pl}\sim {\mathcal O}(10)$
and width parameter $\chi\sim {\mathcal O}(10^{-3})$ constrained by studying inflation \cite{Xue2021} and theoretical calculation \cite{Xue2019,Xue2020}. It shows
the cold dark matter abundance (\ref{cpara3}) can be the right order of 
the magnitude, compared with the currently observed relic value $\Omega^0_c\approx 0.3$.
This preliminary study requires further detailed investigations. Moreover, as CDM candidates, these massive stable pairs should play some role in primordial black hole formulation and primordial gravitational wave emission.

\section{Remarks and summary} 

\subsection{Some remarks}

We study the scenario that the cosmological constant term (dark energy) $\Lambda\not=0$ and $p_{_\Lambda}= -\rho_{_\Lambda}$ represents \cite{Coleman1988, Barvinsky2007,  Xue2009a, Xue2010, Xue2009, Xue2012}
the non-trivial ground state (Wheeler spacetime foam \cite{Misner1973}) 
of the quantum field of spacetime gravity $R$ and $R^2$. Dark energy is neither the vacuum energy of quantum fields of particles (matter) nor the energy of the scalar field's kinetics and potential. We need further investigations to understand $p^{\rm fast, slow}_{_\Lambda}\approx -\rho^{\rm fast, slow}_{_\Lambda}$ of the spacetime foam interacting to matter. 
We study in this article the back-and-forth interactions between dark energy and matter via horizon $H$ in the following two aspects. 

First, at the scale ${\mathcal O}(1/M)$ for fast components and their back-and-forth reactions, we describe in Sec.~\ref{qppo} the quantum massive pair production and oscillation. It forms a quasi-classical coherent state with a large occupation number of particles. 
Averaging fast components over microscopic time, we describe the state as a quasi-classical state of massive pair plasma by using a perfect fluid $p^H_{_M}=\omega^H_{_M}\rho^H_{_M}$ in Sec.~\ref{mpp}. It is functional of dark energy $p_{_\Lambda}^{\rm slow}\approx -\rho_{_\Lambda}^{\rm slow}$ and normal matter $p_{_{R,M}}=\omega_{_{R,M}}\rho_{_{R,M}}$ via the horizon.

Second, at the scale ${\mathcal O}(1/H)$ for slow components and their interactions, we discuss in Sec.~\ref{mppa} the macroscopic 
back-and-forth interacting system between the classical pair plasma fluid $\rho^H_{_M}$, normal matter fluids $\rho_{_{R,M}}$, dark energy $p_{_\Lambda}^{\rm slow}\approx -\rho_{_\Lambda}^{\rm slow}$ and horizon $H_{\rm slow}$ via nonlinear 
Eqs.~(\ref{eqhp}-\ref{reheateq}). The novel cosmic rate equation (\ref{rateeqp}) accounts for the massive pair plasma state interacting with dark energy and normal matter. The system yields $\tilde\Lambda$CDM scenario that describes inflation, reheating and standard cosmology. 

In Ref.~\cite{Xue2021}, we study $\tilde\Lambda$-driven inflation in detail, compare and contrast it with usual canonical inflation models of the scalar field $\phi$, potential $V(\phi)$,
Friedman equation $3m_{\rm pl}^2H^2=\dot\phi^2/2 +V(\phi)$, energy density
$\rho_\phi=\dot\phi^2/2 + V(\phi)$ and pressure $p_\phi=\dot\phi^2/2 - V(\phi)$. The correspondences between inflation models and the $\tilde\Lambda$-driven inflation 
\begin{eqnarray}
\dot\phi^2 \Leftrightarrow \rho^{H}_{_M} 
+ p^{H}_{_M}
\quad 
V(\phi) \Leftrightarrow  \rho_{_\Lambda} + (\rho^{H}_{_M} - p^{H}_{_M})/2,
\label{corr}
\end{eqnarray}
where the dark energy density $\rho_{_{\Lambda}}$ is an $\mathcal{O}(1/H)$ slow component. The $\mathcal{O}(1/M)$ fast oscillating component
$\rho^{\rm fast}_{_{\Lambda}}$ has no relation to the classical field $\phi$. The slow-roll condition  $V(\phi)\gg \dot\phi^2/2$ corresponds 
to $\rho_{_\Lambda} \gg \rho^{H}_{_M}$ for 
$\rho^H_{_M}\approx (2\chi m^2/3m^2_{\rm pl}) \rho_{_\Lambda}$. It leads to $\dot \rho_{_\Lambda} \Leftrightarrow \dot V=\dot\phi V^\prime$ and $\dot \rho^H_{_M} \Leftrightarrow (1/2)d(\dot\phi^2)/dt =\dot\phi\ddot\phi $. 
Equation (\ref{sfriedman}), namely Eq.~(5.1) in Ref.~\cite{Xue2021}, corresponds to the classical equation of motion for $\phi$: $\ddot\phi+3H\dot\phi +V^\prime(\phi)=0$. 
In inflation, we approximately obtain an analytical solution by neglecting the cosmic rate equation (\ref{rateeqd}) because of $\rho^{H}_{_M}\gg \rho_{_{R, M}}$. In reheating, we have to numerically solve the Friedman equations, cosmic rate equation and reheating equation (\ref{eqhp}-\ref{reheateq}), which yield a complex back-and-forth reaction system 
of inter-playing three scales $\tau_{_H}$, $\tau_{_M}$, $\tau_{_D}$ and $\tau_{_R}$ dynamics.

\subsection{Summary}
We make a summary to close this lengthy article. 
In the $\rho_{_\Lambda}$-dominated inflation $H>\Gamma_M$, 
where the massive pair plasma density $\rho^H_{_M}$ is small and normal matter density $\rho_{_M}$ 
is negligible. 
The reheating epoch starts $\Gamma_M> H$, $\rho^H_{_M}$ and $\rho_{_M}$ become large, and their back reaction and decay to relativistic particles are important.
Therefore, the cosmic rate equation (\ref{rateeqd})  
governing the processes $\rho^H_{_M}\Leftrightarrow \rho_{_M}$ (\ref{bfrho}) 
and unstable massive pairs' decay $\bar F F \rightarrow \bar \ell\ell$ are relevant.
It is an additional dynamical equation to two Friedman equations 
(\ref{sfriedman0},\ref{sfriedman}) and the reheating equation (\ref{eqrho}) 
from energy conservation.

Using the massive pair plasma density $\rho^H_{_M}$ 
(\ref{ipe}), production rate $\Gamma_M$ and decay rate 
$\Gamma^{^{\rm de}}_M$ (\ref{rohm}), we study the reheating epoch by a close system of four dynamical (ordinary differential equations) 
equations (\ref{eqhp}-\ref{reheateq}) for the horizon $H$ and three densities $\rho_{_{\Lambda, M,R}}$. The initial conditions are given by the inflation end. Numerically solving this system, we find 
three characteristic episodes:
\begin{enumerate}[(i)]
\item the ${\mathcal P}$-episode of the transition 
from the inflation end to the reheating start, when
the pair-production rate is much larger than the Hubble rate ($\Gamma_M\gg H$), 
the dark energy density $\rho_{_\Lambda}$ quickly decreases and 
converts to
the matter-energy density $\rho_{_M}$. 
As a consequence 
$\rho_{_\Lambda}\ll \rho_{_M}$; 
\item the ${\mathcal M}$-episode of
massive pairs domination, where the back-and-forth interaction 
of the cosmic rate equation (\ref{rateeqd}) 
plays an essential role, and dark energy density slowly varies;
\item the ${\mathcal R}$-episode of the genuine reheating 
$\rho_{_R}\gg \rho_{_M}$, when unstable massive pairs predominately decay to relativistic particles that quickly thermalised. 
\end{enumerate}
We emphasise the pair mass threshold $\hat m> \hat m_{\rm thresh}$ (\ref{mthre}) that at the pre-reheating start 
$\rho_{_\Lambda}\gg \rho_{_M}\gg \rho_{_R}$, 
the rapid converting process 
$\rho_{_\Lambda}\Rightarrow\rho_{_M}\Rightarrow \rho_{_R}$ 
leads to $\rho_{_\Lambda}\ll \rho_{_M}\ll \rho_{_R}$. The most 
relevant mass-energy and entropy 
of Universe are produced by the end of reheating.
Such dynamic processes should lead to the emission of primordial gravitational waves \cite{Maggiore2000, Guzzetti2016}.

The initial conditions for the reheating epoch are the Hubble scale $H_{\rm end}$ and energy densities $\rho^{\rm end}_{_{\Lambda, M}}$ 
at the end of inflation after $e$-folding number $N_{\rm end}$. They are determined by the CMB measurements of scalar amplitude $A_s$ and spectral index $n_s$ at pivot scale $k_*=0.05({\rm Mpc})^{-1}$ 
and the Hubble scale $H_*$ \cite{Xue2021}. We obtain the reheating Hubble 
scale $H_{\rm RH}$, temperature $T_{\rm RH}$ and entropy $S_{\rm patch}$ at 
the genuine reheating episode. 
They are functions of the tensor-to-scalar ratio $r$, and their numerical values are in accordance with the CMB observations. 
Moreover, from purely theoretical viewpoints, we preliminarily limit the $r$ values in the range 
$0< r < 0.047$. 
Among massive pairs gravitationally produced,
(i) unstable pairs decay to relativistic particles, accounting for reheating; (ii) stable pairs couple only to gravity and are candidates for cold dark matter. The resultant cold dark matter abundance $\Omega_c\sim 10^{-1}$ is about a constant in time. They should play the role in the formation of primordial black holes.  

\section{\bf Acknowledgment}
The author thanks the 
EPJC Editor Alexei Starobinsky 
and anonymous referees for their reviews and reports that give chances to improve the manuscript. 

\section{\bf 
Appendix: Quantum pair oscillation details}\label{dis}
In microscopic time, we plot the Bogoliubov coefficient $|\beta|^2$, the quantum pair density $\varrho^{\rm fast}_{_\Lambda}$ and pressure ${\mathcal P}^{\rm fast}_{\Lambda}$, as well as the fast components of Hubble function $h_{\rm fast}$, 
and cosmological term
$\varrho^{\rm fast}_{_\Lambda}$. 
Recall ${\mathcal P}^{\rm fast}_{_\Lambda}\approx -\varrho^{\rm fast}_{_\Lambda}$.

\begin{figure*}[h]
\vspace{+3em}
\includegraphics[height=5.5cm,width=7.8cm]{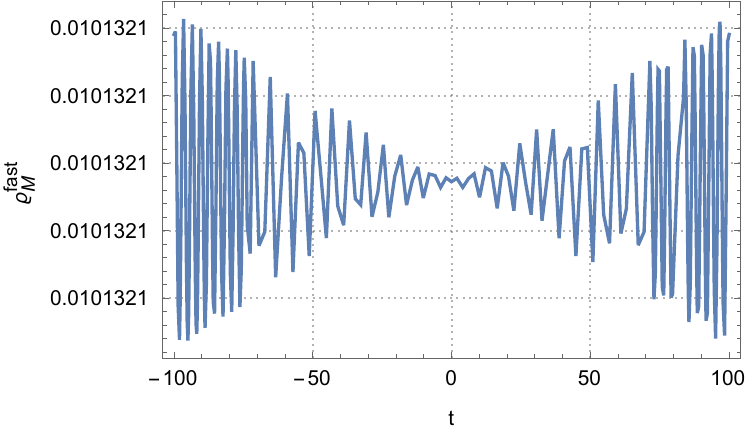}
\hspace{0.333cm}
\includegraphics[height=5.5cm,width=7.8cm]{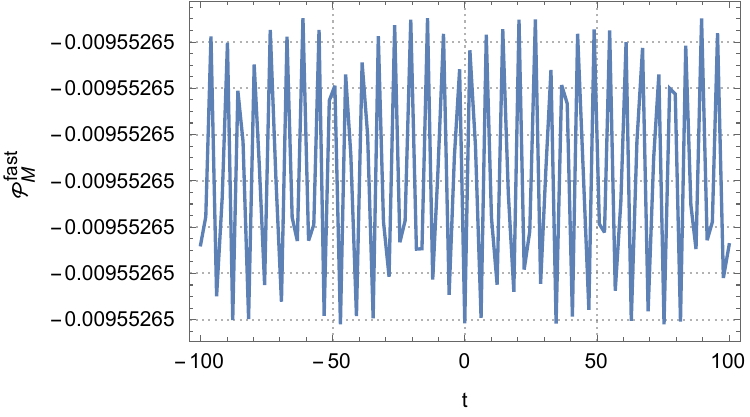}\hspace{0.333cm}
\includegraphics[height=5.5cm,width=7.8cm]{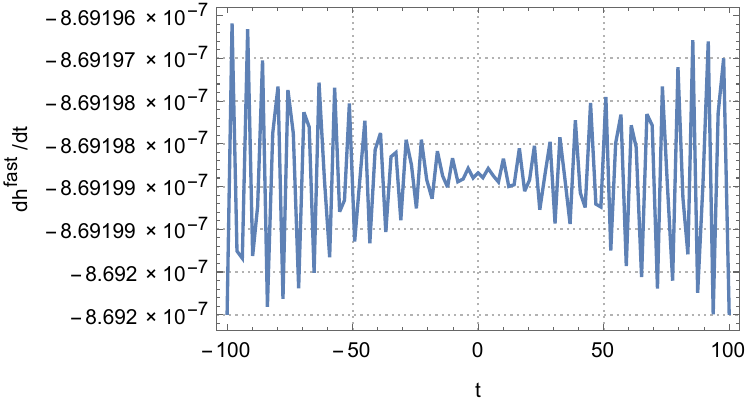}\hspace{0.333cm}
\includegraphics[height=5.5cm,width=7.8cm]{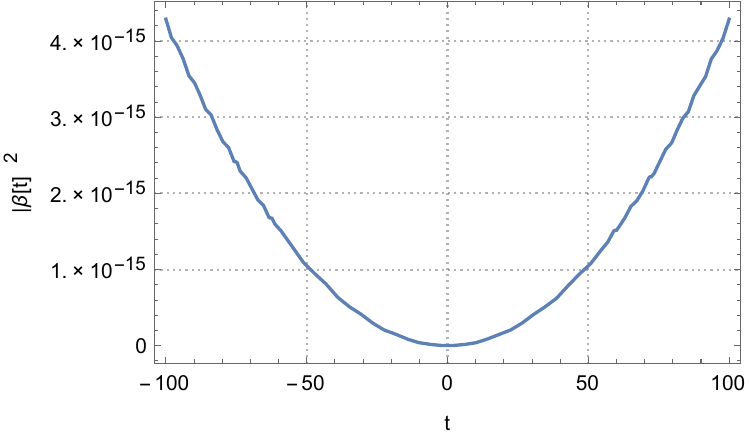}
\includegraphics[height=5.5cm,width=7.8cm]{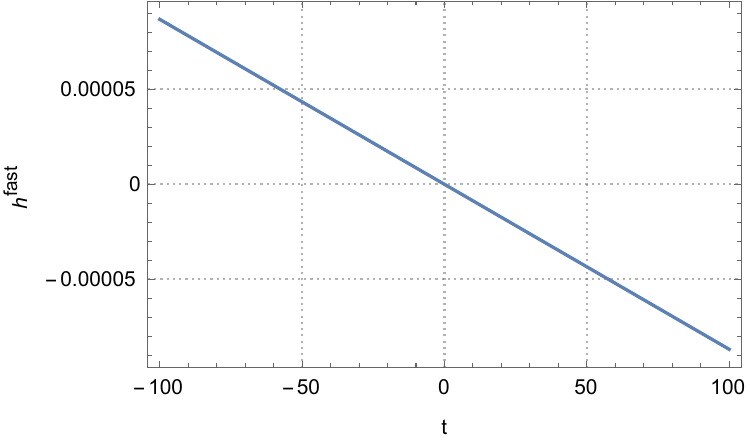}\hspace{0.333cm}
\includegraphics[height=5.5cm,width=7.8cm]{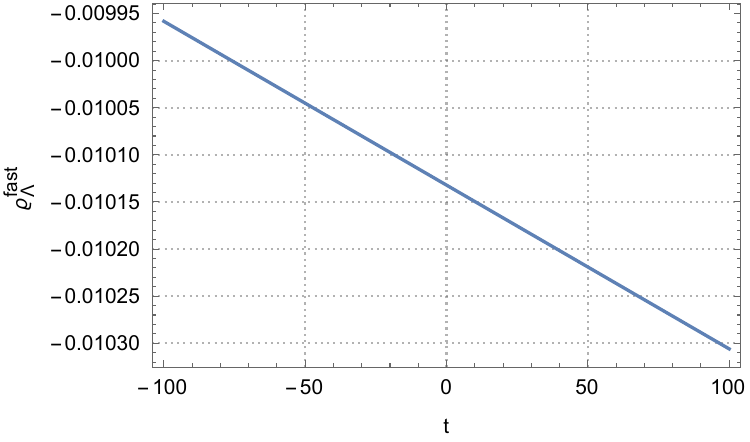}
\caption{Corresponding to Fig.~\ref{reh-osci+}, the details of quantum pair oscillation are shown in microscopic time $t$ in the unit of $M^{-1}$. The parameters' values are the same as those in Fig.~\ref{reh-osci+}. The non-smooth curve $|\beta(t)|^2$ shows its oscillating behavior. The $h_{\rm fast}$ and $\varrho^{\rm fast}_{_\Lambda}$ oscillatory structures are too small to see due to the precision limit for numerical calculations with the parameters' values used. However, one can infer their oscillating behaviours by the oscillating $dh^{\rm fast}/dt$ shown and 
fast-component Eqs.~(\ref{ffriedman+}). We suggest readers see Fig.~4 of Ref.~\cite{Xue2021} for pre-inflation and inflation, where the corresponding solutions for other parameters' values and plotting scales show evident oscillatory structures. The quantum pressure ${\mathcal P}^{\rm fast}_{_M}$ and its time average are negative.}\label{reh-detailosci1+}
\end{figure*}

\providecommand{\href}[2]{#2}\begingroup\raggedright\endgroup

\end{document}